\documentclass{mn2e}
\usepackage{psfig}
\usepackage{mnras_cite}
\def\fileversion{v1.20a}% was \def\fileversion{v1.20}%
\def\filedate{21.6.94}%  was \def\filedate{26.1.94}%
\edef\epsfigRestoreAt{\catcode`@=\number\catcode`@\relax}%
\catcode`\@=11\relax
\ifx\undefined\@makeother                % -pks-
\def\@makeother#1{\catcode`#1=12\relax}  % -pks-
\fi                                      % -pks-
\immediate\write16{Document style option `epsfig', \fileversion\space
<\filedate> (edited by SPQR + pks)}% was <\filedate> (edited by SPQR)}%
\newcount\EPS@Height \newcount\EPS@Width \newcount\EPS@xscale
\newcount\EPS@yscale
\def\psfigdriver#1{%
  \bgroup\edef\next{\def\noexpand\tempa{#1}}%
    \uppercase\expandafter{\next}%
    \def\LN{DVITOLN03}%
    \def\DVItoPS{DVITOPS}%
    \def\DVIPS{DVIPS}%
    \def\emTeX{EMTEX}%
    \def\OzTeX{OZTEX}%
    \def\Textures{TEXTURES}%
    \global\chardef\fig@driver=0
    \ifx\tempa\LN
        \global\chardef\fig@driver=0\fi
    \ifx\tempa\DVItoPS
        \global\chardef\fig@driver=1\fi
    \ifx\tempa\DVIPS
        \global\chardef\fig@driver=2\fi
    \ifx\tempa\emTeX
        \global\chardef\fig@driver=3\fi
    \ifx\tempa\OzTeX
        \global\chardef\fig@driver=4\fi
    \ifx\tempa\Textures
        \global\chardef\fig@driver=5\fi
  \egroup
\def\psfig@start{}%
\def\psfig@end{}%
\def\epsfig@gofer{}%
\ifcase\fig@driver
\typeout{WARNING! ****
 no specials for LN03 psfig}%
\or % case 1: dvitops
\def\psfig@start{}%
\def\psfig@end{\special{dvitops: import \@p@sfilefinal \space
\@p@swidth sp \space \@p@sheight sp \space fill}%
\if@clip \typeout{Clipping not supported}\fi
\if@angle \typeout{Rotating not supported}\fi
}%
\let\epsfig@gofer\psfig@end
\or %case2 dvips
\def\psfig@start{\special{ps::[begin]  \@p@swidth \space \@p@sheight \space%
        \@p@sbbllx \space \@p@sbblly \space%
        \@p@sbburx \space \@p@sbbury \space%
        startTexFig \space }%
        \if@clip
                \if@verbose
                        \typeout{(clipped to BB) }%
                \fi
                \special{ps:: doclip \space }%
        \fi
        \if@angle              % moved after \if@clip ... \fi -pks-
                \special {ps:: \@p@sangle \space rotate \space}
        \fi
        \special{ps: plotfile \@p@sfilefinal \space }%
        \special{ps::[end] endTexFig \space }%
}%
\def\psfig@end{}%
\def\epsfig@gofer{\if@clip
                        \if@verbose
                           \typeout{(clipped to BB)}%
                        \fi
                        \epsfclipon
                  \fi
                  \epsfsetgraph{\@p@sfilefinal}%
}%
\or % case 3, emTeX
\typeout{WARNING. You must have a .bb info file with the Bounding Box
  of the pcx file}%
\def\psfig@start{}%
\def\psfig@end{\typeout{pcx import of \@p@sfilefinal}%
\if@clip \typeout{Clipping not supported}\fi
\if@angle \typeout{Rotating not supported}\fi
\raisebox{\@p@srheight sp}{\special{em: graph \@p@sfilefinal}}}%
\def\epsfig@gofer{}%
\or % case 4, OzTeX
\def\psfig@start{}%
\def\psfig@end{%
\EPS@Width\@p@swidth
\EPS@Height\@p@sheight
\divide\EPS@Width by 65781  % convert sp to bp
\divide\EPS@Height by 65781
\special{epsf=\@p@sfilefinal
\space
width=\the\EPS@Width
\space
height=\the\EPS@Height
}%
\if@clip \typeout{Clipping not supported}\fi
\if@angle \typeout{Rotating not supported}\fi
}%
\let\epsfig@gofer\psfig@end
\or % case 5, Textures
\def\psfig@end{
         \EPS@Width=\@bbw  
         \divide\EPS@Width by 1000
         \EPS@xscale=\@p@swidth \divide \EPS@xscale by \EPS@Width
         \EPS@Height=\@bbh  
         \divide\EPS@Height by 1000
         \EPS@yscale=\@p@sheight \divide \EPS@yscale by\EPS@Height
  \ifnum\EPS@xscale>\EPS@yscale\EPS@xscale=\EPS@yscale\fi
\if@clip
   \if@verbose
      \typeout{(clipped to BB)}%
   \fi
   \epsfclipon
\fi
\special{illustration \@p@sfilefinal\space scaled \the\EPS@xscale}%
}%
\def\psfig@start{}%
\let\epsfig\psfig
\else
\typeout{WARNING. *** unknown  driver - no psfig}%
\fi
}%
\newdimen\ps@dimcent
\ifx\undefined\fbox
\newdimen\fboxrule
\newdimen\fboxsep
\newdimen\ps@tempdima
\newbox\ps@tempboxa
\long\def\fbox#1{\leavevmode\setbox\ps@tempboxa\hbox{#1}\ps@tempdima\fboxrule
    \advance\ps@tempdima \fboxsep \advance\ps@tempdima \dp\ps@tempboxa
   \hbox{\lower \ps@tempdima\hbox
  {\vbox{\hrule height \fboxrule
          \hbox{\vrule width \fboxrule \hskip\fboxsep
          \vbox{\vskip\fboxsep \box\ps@tempboxa\vskip\fboxsep}\hskip
                 \fboxsep\vrule width \fboxrule}%
                 \hrule height \fboxrule}}}}%
\fi
\ifx\@ifundefined\undefined
\long\def\@ifundefined#1#2#3{\expandafter\ifx\csname
  #1\endcsname\relax#2\else#3\fi}%
\fi
\@ifundefined{typeout}%
{\gdef\typeout#1{\immediate\write\sixt@@n{#1}}}%
{\relax}%
\@ifundefined{epsfig}{}{\typeout{EPSFIG --- already loaded} }%
\@ifundefined{epsfbox}{\input epsf}{}%
\ifx\undefined\@latexerr
        \newlinechar`\^^J
        \def\@spaces{\space\space\space\space}%
        \def\@latexerr#1#2{%
        \edef\@tempc{#2}\expandafter\errhelp\expandafter{\@tempc}%
        \typeout{Error. \space see a manual for explanation.^^J
         \space\@spaces\@spaces\@spaces Type \space H <return> \space for
         immediate help.}\errmessage{#1}}%
\fi
\def\@whattodo{You tried to include a PostScript figure which
cannot be found^^JIf you press return to carry on anyway,^^J
The failed name will be printed in place of the figure.^^J
or type X to quit}%
\def\@whattodobb{You tried to include a PostScript figure which
has no^^Jbounding box, and you supplied none.^^J
If you press return to carry on anyway,^^J
The failed name will be printed in place of the figure.^^J
or type X to quit}%
\def\@nnil{\@nil}%
\def\@empty{}%
\def\@psdonoop#1\@@#2#3{}%
\def\@psdo#1:=#2\do#3{\edef\@psdotmp{#2}\ifx\@psdotmp\@empty \else
    \expandafter\@psdoloop#2,\@nil,\@nil\@@#1{#3}\fi}%
\def\@psdoloop#1,#2,#3\@@#4#5{\def#4{#1}\ifx #4\@nnil \else
       #5\def#4{#2}\ifx #4\@nnil \else#5\@ipsdoloop #3\@@#4{#5}\fi\fi}%
\def\@ipsdoloop#1,#2\@@#3#4{\def#3{#1}\ifx #3\@nnil
       \let\@nextwhile=\@psdonoop \else
      #4\relax\let\@nextwhile=\@ipsdoloop\fi\@nextwhile#2\@@#3{#4}}%
\def\@tpsdo#1:=#2\do#3{\xdef\@psdotmp{#2}\ifx\@psdotmp\@empty \else
    \@tpsdoloop#2\@nil\@nil\@@#1{#3}\fi}%
\def\@tpsdoloop#1#2\@@#3#4{\def#3{#1}\ifx #3\@nnil
       \let\@nextwhile=\@psdonoop \else
      #4\relax\let\@nextwhile=\@tpsdoloop\fi\@nextwhile#2\@@#3{#4}}%
\long\def\epsfaux#1#2:#3\\{\ifx#1\epsfpercent
   \def\testit{#2}\ifx\testit\epsfbblit
        \@atendfalse
        \epsf@atend #3 . \\%
        \if@atend
           \if@verbose
                \typeout{epsfig: found `(atend)'; continuing search}%
           \fi
        \else
                \epsfgrab #3 . . . \\%
                \epsffileokfalse\global\no@bbfalse
                \global\epsfbbfoundtrue
        \fi
   \fi\fi}%
\def\epsf@atendlit{(atend)}
\def\epsf@atend #1 #2 #3\\{%
   \def\epsf@tmp{#1}\ifx\epsf@tmp\empty
      \epsf@atend #2 #3 .\\\else
   \ifx\epsf@tmp\epsf@atendlit\@atendtrue\fi\fi}%

\chardef\trig@letter = 11
\chardef\other = 12
 
\newif\ifdebug %%% turn me on to see TeX hard at work ...
\newif\ifc@mpute %%% don't need to compute some values
\newif\if@atend
\c@mputetrue % but assume that we do
 
\let\then = \relax
\def\r@dian{pt }%
\let\r@dians = \r@dian
\let\dimensionless@nit = \r@dian
\let\dimensionless@nits = \dimensionless@nit
\def\internal@nit{sp }%
\let\internal@nits = \internal@nit
\newif\ifstillc@nverging
\def \Mess@ge #1{\ifdebug \then \message {#1} \fi}%
 
{ %%% Things that need abnormal catcodes %%%
        \catcode `\@ = \trig@letter
        \gdef \nodimen {\expandafter \n@dimen \the \dimen}%
        \gdef \term #1 #2 #3%
               {\edef \t@ {\the #1}%%% freeze parameter 1 (count, by value)
                \edef \t@@ {\expandafter \n@dimen \the #2\r@dian}%
                \t@rm {\t@} {\t@@} {#3}%
               }%
        \gdef \t@rm #1 #2 #3%
               {{%
                \count 0 = 0
                \dimen 0 = 1 \dimensionless@nit
                \dimen 2 = #2\relax
                \Mess@ge {Calculating term #1 of \nodimen 2}%
                \loop
                \ifnum  \count 0 < #1
                \then   \advance \count 0 by 1
                        \Mess@ge {Iteration \the \count 0 \space}%
                        \Multiply \dimen 0 by {\dimen 2}%
                        \Mess@ge {After multiplication, term = \nodimen 0}%
                        \Divide \dimen 0 by {\count 0}%
                        \Mess@ge {After division, term = \nodimen 0}%
                \repeat
                \Mess@ge {Final value for term #1 of
                                \nodimen 2 \space is \nodimen 0}%
                \xdef \Term {#3 = \nodimen 0 \r@dians}%
                \aftergroup \Term
               }}%
        \catcode `\p = \other
        \catcode `\t = \other
        \gdef \n@dimen #1pt{#1} %%% throw away the ``pt''
}%
 
\def \Divide #1by #2{\divide #1 by #2} %%% just a synonym
 
\def \Multiply #1by #2%%% allows division of a dimen by a dimen
       {{%%% should really freeze parameter 2 (dimen, passed by value)
        \count 0 = #1\relax
        \count 2 = #2\relax
        \count 4 = 65536
        \Mess@ge {Before scaling, count 0 = \the \count 0 \space and
                        count 2 = \the \count 2}%
        \ifnum  \count 0 > 32767 %%% do our best to avoid overflow
        \then   \divide \count 0 by 4
                \divide \count 4 by 4
        \else   \ifnum  \count 0 < -32767
                \then   \divide \count 0 by 4
                        \divide \count 4 by 4
                \else
                \fi
        \fi
        \ifnum  \count 2 > 32767 %%% while retaining reasonable accuracy
        \then   \divide \count 2 by 4
                \divide \count 4 by 4
        \else   \ifnum  \count 2 < -32767
                \then   \divide \count 2 by 4
                        \divide \count 4 by 4
                \else
                \fi
        \fi
        \multiply \count 0 by \count 2
        \divide \count 0 by \count 4
        \xdef \product {#1 = \the \count 0 \internal@nits}%
        \aftergroup \product
       }}%
 
\def\r@duce{\ifdim\dimen0 > 90\r@dian \then   % sin(x) = sin(180-x)
                \multiply\dimen0 by -1
                \advance\dimen0 by 180\r@dian
                \r@duce
            \else \ifdim\dimen0 < -90\r@dian \then  % sin(x) = sin(360+x)
                \advance\dimen0 by 360\r@dian
                \r@duce
                \fi
            \fi}%
 
\def\Sine#1%
       {{%
        \dimen 0 = #1 \r@dian
        \r@duce
        \ifdim\dimen0 = -90\r@dian \then
           \dimen4 = -1\r@dian
           \c@mputefalse
        \fi
        \ifdim\dimen0 = 90\r@dian \then
           \dimen4 = 1\r@dian
           \c@mputefalse
        \fi
        \ifdim\dimen0 = 0\r@dian \then
           \dimen4 = 0\r@dian
           \c@mputefalse
        \fi
        \ifc@mpute \then
                \divide\dimen0 by 180
                \dimen0=3.141592654\dimen0
                \dimen 2 = 3.1415926535897963\r@dian %%% a well-known constant
                \divide\dimen 2 by 2 %%% we only deal with -pi/2 : pi/2
                \Mess@ge {Sin: calculating Sin of \nodimen 0}%
                \count 0 = 1 %%% see power-series expansion for sine
                \dimen 2 = 1 \r@dian %%% ditto
                \dimen 4 = 0 \r@dian %%% ditto
                \loop
                        \ifnum  \dimen 2 = 0 %%% then we've done
                        \then   \stillc@nvergingfalse
                        \else   \stillc@nvergingtrue
                        \fi
                        \ifstillc@nverging %%% then calculate next term
                        \then   \term {\count 0} {\dimen 0} {\dimen 2}%
                                \advance \count 0 by 2
                                \count 2 = \count 0
                                \divide \count 2 by 2
                                \ifodd  \count 2 %%% signs alternate
                                \then   \advance \dimen 4 by \dimen 2
                                \else   \advance \dimen 4 by -\dimen 2
                                \fi
                \repeat
        \fi
                        \xdef \sine {\nodimen 4}%
       }}%
 
\def\Cosine#1{\ifx\sine\UnDefined\edef\Savesine{\relax}\else
                             \edef\Savesine{\sine}\fi
        {\dimen0=#1\r@dian\multiply\dimen0 by -1
         \advance\dimen0 by 90\r@dian
         \Sine{\nodimen 0}%
         \xdef\cosine{\sine}%
         \xdef\sine{\Savesine}}}
\def\psdraft{\def\@psdraft{0}}%
\def\psfull{\def\@psdraft{1}}%
\psfull
\newif\if@compress
\def\pscompress{\@compresstrue}
\def\psnocompress{\@compressfalse}
\@compressfalse
\newif\if@scalefirst
\def\psscalefirst{\@scalefirsttrue}%
\def\psrotatefirst{\@scalefirstfalse}%
\psrotatefirst
\newif\if@draftbox
\def\psnodraftbox{\@draftboxfalse}%
\@draftboxtrue
\newif\if@noisy
\@noisyfalse
\newif\ifno@bb
\newif\if@bbllx
\newif\if@bblly
\newif\if@bburx
\newif\if@bbury
\newif\if@height
\newif\if@width
\newif\if@rheight
\newif\if@rwidth
\newif\if@angle
\newif\if@clip
\newif\if@verbose
\newif\if@prologfile
\def\@p@@sprolog#1{\@prologfiletrue\def\@prologfileval{#1}}%
\def\@p@@sclip#1{\@cliptrue}%
\newif\ifepsfig@dos  % only single suffix possible
\def\epsfigdos{\epsfig@dostrue}%
\epsfig@dosfalse
\newif\ifuse@psfig
\def\ParseName#1{\expandafter\@Parse#1}%
\def\@Parse#1.#2:{\gdef\BaseName{#1}\gdef\FileType{#2}}%

\def\@p@@sfile#1{%
  \ifepsfig@dos
     \ParseName{#1:}%
  \else
     \gdef\BaseName{#1}\gdef\FileType{}%
  \fi
  \def\@p@sfile{NO FILE: #1}%
  \def\@p@sfilefinal{NO FILE: #1}%
  \openin1=#1
  \ifeof1\closein1\openin1=\BaseName.bb
    \ifeof1\closein1
      \if@bbllx                 % No postscript file but bb given explicitly.
        \if@bblly\if@bburx\if@bbury
          \def\@p@sfile{#1}%
          \def\@p@sfilefinal{#1}%
        \fi\fi\fi
      \else                     % No bounding box found.
        \@latexerr{ERROR. PostScript file #1 not found}\@whattodo
        \@p@@sbbllx{100bp}%
        \@p@@sbblly{100bp}%
        \@p@@sbburx{200bp}%
        \@p@@sbbury{200bp}%
        \psdraft
      \fi
    \else                       % Postscript file is compressed.
      \closein1%
      \edef\@p@sfile{\BaseName.bb}%
      \typeout{using BB from \@p@sfile}%
      \ifnum\fig@driver=3
        \edef\@p@sfilefinal{\BaseName.pcx}%
      \else
        \ifepsfig@dos
          \edef\@p@sfilefinal{"`gunzip -c `texfind \BaseName.{z,Z,gz}"}%
        \else
          \edef\@p@sfilefinal{"`epsfig \if@compress-c \fi#1"}%          
        \fi
      \fi
    \fi
  \else\closein1                % Postscript file is not compressed.
    \edef\@p@sfile{#1}%
    \if@compress  
      \edef\@p@sfilefinal{"`epsfig -c #1"}%
    \else
      \edef\@p@sfilefinal{#1}%
    \fi
  \fi%
}

\let\@p@@sfigure\@p@@sfile
\def\@p@@sbbllx#1{%
                                            \@bbllxtrue
                \ps@dimcent=#1
                \edef\@p@sbbllx{\number\ps@dimcent}%
                \divide\ps@dimcent by65536
                \global\edef\epsfllx{\number\ps@dimcent}%
}%
\def\@p@@sbblly#1{%
                \@bbllytrue
                \ps@dimcent=#1
                \edef\@p@sbblly{\number\ps@dimcent}%
                \divide\ps@dimcent by65536
                \global\edef\epsflly{\number\ps@dimcent}%
}%
\def\@p@@sbburx#1{%
                \@bburxtrue
                \ps@dimcent=#1
                \edef\@p@sbburx{\number\ps@dimcent}%
                \divide\ps@dimcent by65536
                \global\edef\epsfurx{\number\ps@dimcent}%
}%
\def\@p@@sbbury#1{%
                \@bburytrue
                \ps@dimcent=#1
                \edef\@p@sbbury{\number\ps@dimcent}%
                \divide\ps@dimcent by65536
                \global\edef\epsfury{\number\ps@dimcent}%
}%
\def\@p@@sheight#1{%
                \@heighttrue
                \global\epsfysize=#1
                \ps@dimcent=#1
                \edef\@p@sheight{\number\ps@dimcent}%
}%
\def\@p@@swidth#1{%
                \@widthtrue
                \global\epsfxsize=#1
                \ps@dimcent=#1
                \edef\@p@swidth{\number\ps@dimcent}% 
}%
\def\@p@@srheight#1{%
                \@rheighttrue\use@psfigtrue
                \ps@dimcent=#1
                \edef\@p@srheight{\number\ps@dimcent}%
}%
\def\@p@@srwidth#1{%
                \@rwidthtrue\use@psfigtrue
                \ps@dimcent=#1
                \edef\@p@srwidth{\number\ps@dimcent}%
}%
\def\@p@@sangle#1{%
                \use@psfigtrue
                \@angletrue
                \edef\@p@sangle{#1}%
}%
\def\@p@@ssilent#1{%
                \@verbosefalse
}%
\def\@p@@snoisy#1{%
                \@verbosetrue
}%
\def\@cs@name#1{\csname #1\endcsname}%
\def\@setparms#1=#2,{\@cs@name{@p@@s#1}{#2}}%
\def\ps@init@parms{%
                \@bbllxfalse \@bbllyfalse
                \@bburxfalse \@bburyfalse
                \@heightfalse \@widthfalse
                \@rheightfalse \@rwidthfalse
                \def\@p@sbbllx{}\def\@p@sbblly{}%
                \def\@p@sbburx{}\def\@p@sbbury{}%
                \def\@p@sheight{}\def\@p@swidth{}%
                \def\@p@srheight{}\def\@p@srwidth{}%
                \def\@p@sangle{0}%
                \def\@p@sfile{}%
                \use@psfigfalse
                \@prologfilefalse
                \def\@sc{}%
                \if@noisy
                        \@verbosetrue
                \else
                        \@verbosefalse
                \fi
                \@clipfalse
}%
\def\parse@ps@parms#1{%
                \@psdo\@psfiga:=#1\do
                   {\expandafter\@setparms\@psfiga,}%
\if@prologfile
\fi
}%
\def\bb@missing{%
        \if@verbose
            \typeout{psfig: searching \@p@sfile \space  for bounding box}%
        \fi
        \epsfgetbb{\@p@sfile}%
        \ifepsfbbfound
            \ps@dimcent=\epsfllx bp\edef\@p@sbbllx{\number\ps@dimcent}%
            \ps@dimcent=\epsflly bp\edef\@p@sbblly{\number\ps@dimcent}%
            \ps@dimcent=\epsfurx bp\edef\@p@sbburx{\number\ps@dimcent}%
            \ps@dimcent=\epsfury bp\edef\@p@sbbury{\number\ps@dimcent}%
        \else
            \epsfbbfoundfalse
        \fi
}
\newdimen\p@intvaluex
\newdimen\p@intvaluey
\def\rotate@#1#2{{\dimen0=#1 sp\dimen1=#2 sp
                  \global\p@intvaluex=\cosine\dimen0
                  \dimen3=\sine\dimen1
                  \global\advance\p@intvaluex by -\dimen3
                  \global\p@intvaluey=\sine\dimen0
                  \dimen3=\cosine\dimen1
                  \global\advance\p@intvaluey by \dimen3
                  }}%
\def\compute@bb{%
                \epsfbbfoundfalse
                \if@bbllx\epsfbbfoundtrue\fi
                \if@bblly\epsfbbfoundtrue\fi
                \if@bburx\epsfbbfoundtrue\fi
                \if@bbury\epsfbbfoundtrue\fi
                \ifepsfbbfound\else\bb@missing\fi
                \ifepsfbbfound\else
                \@latexerr{ERROR. cannot locate BoundingBox}\@whattodobb
                        \@p@@sbbllx{100bp}%
                        \@p@@sbblly{100bp}%
                        \@p@@sbburx{200bp}%
                        \@p@@sbbury{200bp}%
                        \no@bbtrue
                        \psdraft
                \fi
                \count203=\@p@sbburx
                \count204=\@p@sbbury
                \advance\count203 by -\@p@sbbllx
                \advance\count204 by -\@p@sbblly
                \edef\ps@bbw{\number\count203}%
                \edef\ps@bbh{\number\count204}%
                 \edef\@bbw{\number\count203}%
                \edef\@bbh{\number\count204}%
               \if@angle
                        \Sine{\@p@sangle}\Cosine{\@p@sangle}%
 
{\ps@dimcent=\maxdimen\xdef\r@p@sbbllx{\number\ps@dimcent}%
 
\xdef\r@p@sbblly{\number\ps@dimcent}%
 
\xdef\r@p@sbburx{-\number\ps@dimcent}%
 
\xdef\r@p@sbbury{-\number\ps@dimcent}}%
                        \def\minmaxtest{%
                           \ifnum\number\p@intvaluex<\r@p@sbbllx
                              \xdef\r@p@sbbllx{\number\p@intvaluex}\fi
                           \ifnum\number\p@intvaluex>\r@p@sbburx
                              \xdef\r@p@sbburx{\number\p@intvaluex}\fi
                           \ifnum\number\p@intvaluey<\r@p@sbblly
                              \xdef\r@p@sbblly{\number\p@intvaluey}\fi
                           \ifnum\number\p@intvaluey>\r@p@sbbury
                              \xdef\r@p@sbbury{\number\p@intvaluey}\fi
                           }%
                        \rotate@{\@p@sbbllx}{\@p@sbblly}%
                        \minmaxtest
                        \rotate@{\@p@sbbllx}{\@p@sbbury}%
                        \minmaxtest
                        \rotate@{\@p@sbburx}{\@p@sbblly}%
                        \minmaxtest
                        \rotate@{\@p@sbburx}{\@p@sbbury}%
                        \minmaxtest
 
\edef\@p@sbbllx{\r@p@sbbllx}\edef\@p@sbblly{\r@p@sbblly}%
 
\edef\@p@sbburx{\r@p@sbburx}\edef\@p@sbbury{\r@p@sbbury}%
                \fi
                \count203=\@p@sbburx
                \count204=\@p@sbbury
                \advance\count203 by -\@p@sbbllx
                \advance\count204 by -\@p@sbblly
                \edef\@bbw{\number\count203}%
                \edef\@bbh{\number\count204}%
}%
\def\in@hundreds#1#2#3{\count240=#2 \count241=#3
                     \count100=\count240        % 100 is first digit #2/#3
                     \divide\count100 by \count241
                     \count101=\count100
                     \multiply\count101 by \count241
                     \advance\count240 by -\count101
                     \multiply\count240 by 10
                     \count101=\count240        %101 is second digit of #2/#3
                     \divide\count101 by \count241
                     \count102=\count101
                     \multiply\count102 by \count241
                     \advance\count240 by -\count102
                     \multiply\count240 by 10
                     \count102=\count240        % 102 is the third digit
                     \divide\count102 by \count241
                     \count200=#1\count205=0
                     \count201=\count200
                        \multiply\count201 by \count100
                        \advance\count205 by \count201
                     \count201=\count200
                        \divide\count201 by 10
                        \multiply\count201 by \count101
                        \advance\count205 by \count201
                     \count201=\count200
                        \divide\count201 by 100
                        \multiply\count201 by \count102
                        \advance\count205 by \count201
                     \edef\@result{\number\count205}%
}%
\def\compute@wfromh{%
                \in@hundreds{\@p@sheight}{\@bbw}{\@bbh}%
                \edef\@p@swidth{\@result}%
}%
\def\compute@hfromw{%
                \in@hundreds{\@p@swidth}{\@bbh}{\@bbw}%
                \edef\@p@sheight{\@result}%
}%
\def\compute@handw{%
                \if@height
                        \if@width
                        \else
                                \compute@wfromh
                        \fi
                \else
                        \if@width
                                \compute@hfromw
                        \else
                                \edef\@p@sheight{\@bbh}%
                                \edef\@p@swidth{\@bbw}%
                        \fi
                \fi
}%
\def\compute@resv{%
                \if@rheight \else \edef\@p@srheight{\@p@sheight} \fi
                \if@rwidth \else \edef\@p@srwidth{\@p@swidth} \fi
}%
\def\compute@sizes{%
        \if@scalefirst\if@angle
        \if@width
           \in@hundreds{\@p@swidth}{\@bbw}{\ps@bbw}%
           \edef\@p@swidth{\@result}%
        \fi
        \if@height
           \in@hundreds{\@p@sheight}{\@bbh}{\ps@bbh}%
           \edef\@p@sheight{\@result}%
        \fi
        \fi\fi
        \compute@handw
        \compute@resv
}
\long\def\graphic@verb#1{\def\next{#1}%
  {\expandafter\graphic@strip\meaning\next}}
\def\graphic@strip#1>{}
\def\graphic@zapspace#1{%
  #1\ifx\graphic@zapspace#1\graphic@zapspace%
  \else\expandafter\graphic@zapspace%
  \fi}
\def\psfig#1{%
\edef\@tempa{\graphic@zapspace#1{}}%
\ifvmode\leavevmode\fi\vbox {%
        \ps@init@parms
        \parse@ps@parms{\@tempa}%
        \ifnum\@psdraft=1
                \typeout{[\@p@sfilefinal]}%
                \if@verbose
                        \typeout{epsfig: using PSFIG macros}%
                \fi
                \psfig@method
        \else
                \epsfig@draft
        \fi
}
}%
\def\graphic@zapspace#1{%
  #1\ifx\graphic@zapspace#1\graphic@zapspace%
  \else\expandafter\graphic@zapspace%
  \fi}
\def\epsfig#1{%
\edef\@tempa{\graphic@zapspace#1{}}%
\ifvmode\leavevmode\fi\vbox {%
        \ps@init@parms
        \parse@ps@parms{\@tempa}%
        \ifnum\@psdraft=1
          \if@angle\use@psfigtrue\fi
          {\ifnum\fig@driver=1\global\use@psfigtrue\fi}%
          {\ifnum\fig@driver=3\global\use@psfigtrue\fi}%
          {\ifnum\fig@driver=4\global\use@psfigtrue\fi}%
          {\ifnum\fig@driver=5\global\use@psfigtrue\fi}%
                \ifuse@psfig
                        \if@verbose
                                \typeout{epsfig: using PSFIG macros}%
                        \fi
                        \psfig@method
                \else
                        \if@verbose
                                \typeout{epsfig: using EPSF macros}%
                        \fi
                        \epsf@method
                \fi
        \else
                \epsfig@draft
        \fi
}%
}%

\def\epsf@method{%
        \epsfbbfoundfalse
        \if@bbllx\epsfbbfoundtrue\fi
        \if@bblly\epsfbbfoundtrue\fi
        \if@bburx\epsfbbfoundtrue\fi
        \if@bbury\epsfbbfoundtrue\fi
        \ifepsfbbfound\else\epsfgetbb{\@p@sfile}\fi
        \ifepsfbbfound
           \typeout{<\@p@sfilefinal>}%
           \epsfig@gofer
        \else
          \@latexerr{ERROR - Cannot locate BoundingBox}\@whattodobb
          \@p@@sbbllx{100bp}%
          \@p@@sbblly{100bp}%
          \@p@@sbburx{200bp}%
          \@p@@sbbury{200bp}%
                \count203=\@p@sbburx
                \count204=\@p@sbbury
                \advance\count203 by -\@p@sbbllx
                \advance\count204 by -\@p@sbblly
                \edef\@bbw{\number\count203}%
                \edef\@bbh{\number\count204}%
          \compute@sizes
          \epsfig@@draft
       \fi
}%
\def\psfig@method{%
        \compute@bb
        \ifepsfbbfound
          \compute@sizes
          \psfig@start
          \vbox to \@p@srheight sp{\hbox to \@p@srwidth 
            sp{\hss}\vss\psfig@end}%
        \else
           \epsfig@draft
        \fi
}%
\def\epsfig@draft{\compute@bb\compute@sizes\epsfig@@draft}%
\def\epsfig@@draft{%
\typeout{<(draft only) \@p@sfilefinal>}%
\if@draftbox
        \hbox{{\fboxsep0pt\fbox{\vbox to \@p@srheight sp{%
        \vss\hbox to \@p@srwidth sp{ \hss 
           \expandafter\Literally\@p@sfilefinal\@nil
                          \hss }\vss
        }}}}%
\else
        \vbox to \@p@srheight sp{%
        \vss\hbox to \@p@srwidth sp{\hss}\vss}%
\fi
}%
\def\Literally#1\@nil{{\tt\graphic@verb{#1}}}
\psfigdriver{dvips}%
\epsfigRestoreAt

\begin{document}

\def \vjec{\vfill\eject}
\def\kmsmpc{\mathrm{km}\ \mathrm{s^{-1}} \mathrm{Mpc^{-1}}}
\def\hmpc{h^{-1} \mathrm{Mpc}}
\def\hkpc{h^{-1} \mathrm{kpc}}
\def\msun{M_\odot}
\def\hmsun{h^{-1} M_\odot}
 \def\chisq{$\chi^2$}
\def \m3{{\rm Mark III}}
\def \etal {{et al.\ }}
\def \cf {{\it cf.\ }}
\def \vs {{\it vs.\ }}
\def \via {{\it via\ }}
\def \ie {{\it i.e.\ } }
\def \eg{{\it e.g.\ }}
\def\kms{\mathrm{km~s^{-1}}}
\def\br{{\bf r}}
\def\bv{{\bf v}}
\def\bV{{\bf V}}

\def\plotfour#1#2#3#4{\centering \leavevmode
\epsfxsize=.24\columnwidth \epsfbox{#1} \hfil
\epsfxsize=.24\columnwidth \epsfbox{#2} \hfil
\epsfxsize=.24\columnwidth \epsfbox{#3} \hfil
\epsfxsize=.24\columnwidth \epsfbox{#4}}

\def\gsim{~\rlap{$>$}{\lower 1.0ex\hbox{$\sim$}}}
\def\lsim{~\rlap{$<$}{\lower 1.0ex\hbox{$\sim$}}}
\def\d{{\rm d}}
\def\c{{\rm c}}
\def \<{\langle}
\def \>{\rangle}

\newcommand{\pa}{\partial}
\newcommand{\cri}{_{\rm cr}}

\title[Two Body Relaxation in Simulated Cosmological Haloes]
{Two Body Relaxation in Simulated Cosmological Haloes}

\author[A. A.~El-Zant]
{Amr A. El-Zant \\ 
CITA, University of Toronto. Ontario M5S 3H8, Canada}

\maketitle

\begin{abstract}
This paper aims at quantifying discreetness effects, born of finite particle number,
on the dynamics of dark matter haloes  
forming in the context of cosmological simulations. By generalising the standard calculation of
two body relaxation to the case when the size and mass distribution are variable,   
and parametrising the time  evolution  using  established empirical relations,   
we find that the dynamics of a million particle  halo is noise-dominated within the inner 
percent of the final virial radius. Far larger particle numbers ($\sim 10^{8}$) are required 
for the RMS perturbations to the velocity to drop to the $10 \%$ level there.
The  radial scaling of the relaxation time is simple and strong:\\ 
$t_{\rm relax} \sim r^2$, 
implying that numbers $\gg 10^8$ are required to faithfully model the very inner regions; 
artificial relaxation may thus constitute an important factor, contributing to the contradictory 
claims concerning the persistence of a power law density cusp to the very centre.
 The cores of substructure haloes can be many relaxation times old. 
Since relaxation first causes their expansion before recontraction occurs, 
it may  render  them  either  more difficult or easier to disrupt, depending
on their orbital parameters. It  may thus modify the  characteristics of the  subhalo distribution; and, 
if as suggested by several authors, it is parent-satellite interactions that determine halo profiles,   
the overall structure of the system may be affected. We  derive simple closed form 
formulas for the characteristic relaxation time of both parents and satellites, and 
an elementary argument deducing the weak $N$-scaling reported by Diemand et al. (2004) 
when the main contribution comes from relaxing subhaloes.

\end{abstract}


%%%%%%%%%%%%%%%%%%%%%%%%%%%%%%%%%%

\section{Introduction}
\label{sec:intro}

 $N$-body  modelling of the dynamical evolution of the dark matter 
component involves the approximation whereby the system to be simulated
is represented by a surrogate one consisting of many order of 
magnitudes fewer particles. Discreteness noise is thus necessarilly greatly
enhanced. Furthermore, within the context of the  cold dark matter scenario,
material rapidly collapses into haloes; in a hierarchical process that ensures
that the first structures are badly resolved, with particle noise 
propagating through the hierarchy, further enhancing discreteness effects
and  weakening the $N$-dependence of the effective relaxation time 
(Binney \& Knebe 2002; Diemand et al. 2004). 

Relaxation is expected to play its most prominent role in the very central regions of collapsed structures  --- 
precisely the place where there remains considerable controversy as to whether the slope of the  
density profile persists as  a power law up to the resolution limit (Diemand et al. 2005), or instead 
gradually flattens into a smooth core (Navarro et al. 2004). From the difference between these two situations
follow important implications concerning the compatibility of the cold dark matter scenario with the observed 
inner mass distribution in galaxies. Relaxation  also  affects the symmetry of a system 
in physical and velocity space, rendering it more isotropic; and triaxial  haloes  may 
play an important role in determining the dynamics of the baryonic component 
(e.g., El-Zant \& Shlosman 2002; El-Zant et al. 2003).

Since substructure haloes  are resolved with far fewer particles, relaxation is expected to have a 
more dominant effect on their internal structure and (by modifying the efficiency of stripping) 
their spatial distribution. If accretion and merging are
are dominant processes in determining them (\pcite{sywt98,dekev03,dekar03}), 
this  will have important  implications for the parent halo properties too.

Estimating the importance of particle noise in $N$-body simulations is an 
essential step in evaluating the role it may play in determining  
the dynamical state of collapsed structures.
Our focus here is on the generalisation of the standard formulation of the problem, leading to the 
familiar formulas for the `relaxation time' in stellar systems, to the cosmologically relevant case, 
when the physical extent and the mass distribution of the object under consideration continuously vary with time. 
The details of the procedure, along with the assumptions involved, are outlined in Section~\ref{sec:relax}. In our treatment we will distinguish
between `parent' haloes, representing the most massive progenitors and `subhaloes', representing the 
substructure. The former are assumed to continue to grow in mass and size, while the latter's
growth stops once they are incorporated into a larger structure. In both cases we derive 
simple formulas for calculating characteristic relaxation times (Section~\ref{sec:simple}) and, in addition, 
undertake a more detailed calculation of the expected {\em root mean square (RMS)} perturbation to trajectories' velocities
for haloes  evolving in the context of the `concordance' CDM 
model; giving simple empirical fits to the results (Section~\ref{sec:full}). 
Conclusions are summarised in Section~\ref{sec:conc}.

\section{The model}
\label{sec:relax}

\begin{figure*}
\psfig{file=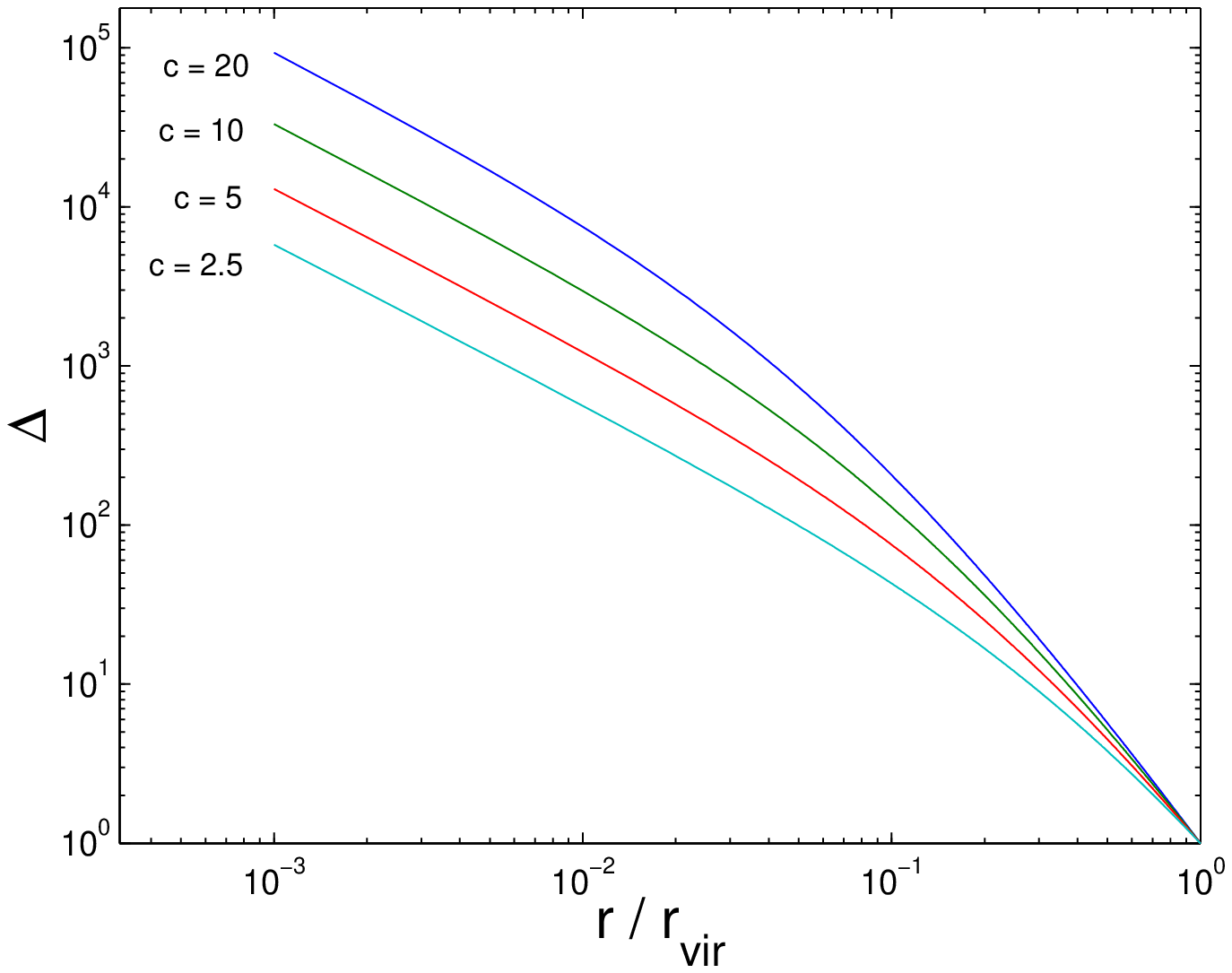,width=87mm}
\psfig{file=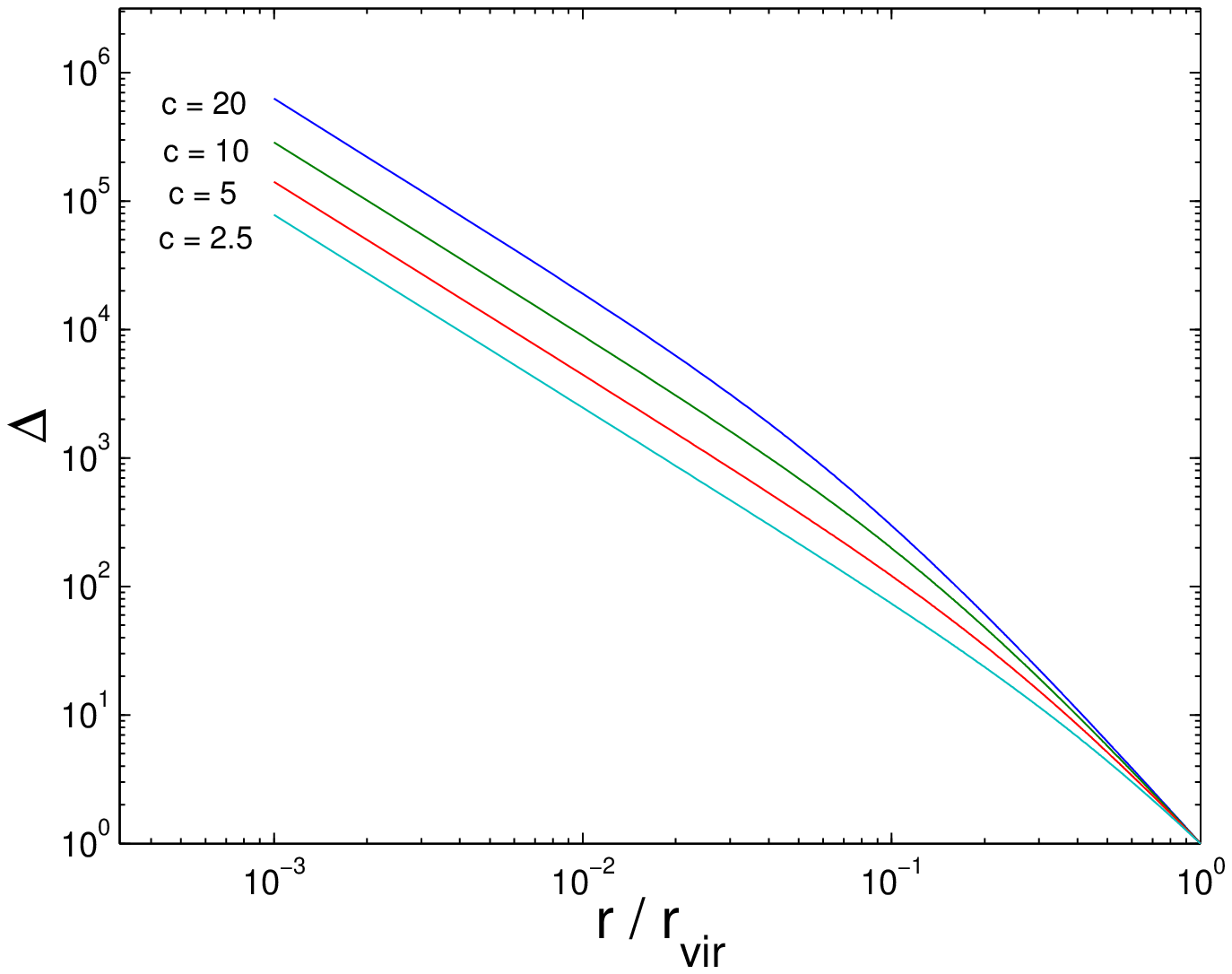,width=87mm}
%\onecolumn
\caption{Density contrast $\Delta$, measured relative to the average density 
within the virial radius, for NFW (left) and M99 profiles 
with different values for the concentration parameters $c$}
\label{fig:overdens}
\end{figure*}

We are interested in the description of relaxation due to discreteness
(particle) noise in evolving cosmological haloes. The model through which 
this is done is described below. First the time  evolution of the halo mass 
distribution is parametrised; the standard description 
of relaxation in gravitational systems is then generalised 
to encompass the case of these evolving systems.

\subsection{Evolution of the halo mass distribution}
\label{sec:massmod}

Describing the temporal evolution of the halo mass 
distribution is simplified by the fact that their densities can  
always be approximately  parametrized by double power law functions, 
independently  of the mass of 
the halo, or the redshift at which it is identified. 
Two profiles have been extensively used: that introduced by~\pcite{nfw} (NFW)
 has density $\sim 1/r$ in the inner region and $\sim 1/r^3$ in the outer one;
the  associated mass fraction within radius $r$ is 
\begin{equation}
\frac{M(r)}{M (r_{\rm vir})} = 
 \frac{\ln (1 + r/r_s) - \frac{r/r_s}{1 + r/r_s}}{\ln (1 + c) - \frac{c}{1 + c}}.
\label{eq:massnfw}
\end{equation}
The profile put forward by~\pcite{m99} (M99)  has similar asymptotic density
variation at large $r$, but a steeper central cusp ($\sim 1/r^{3/2}$). It has mass 
distribution
\begin{equation}
\frac{M(r)}{M (r_{\rm vir})} = \frac{\ln [1 + (r/r_s)^{3/2}]}{\ln \left(1 + c^{3/2}\right)}.
\label{eq:massmoore}
\end{equation}
In these formulas, $r_s$ is the characteristic halo scale length (separating the $1/r$ from the 
$1/r^3$ region) and $r_{\rm vir}$
is the virial radius. Their ratio $c= r_{\rm vir}/ r_s$ determines the 
mean density contrast inside  radius $r$:
$\Delta = \frac{M(r)}{ M(r_{\rm vir}) } \frac{r^3_{\rm vir}}{r^3}$ (cf. Fig~\ref{fig:overdens}).

The virial radius  is usually  defined as the radius
inside which the average density is $\sim 200$ times the critical density for 
closure $\rho_c$. Relative to $\rho_c$, the   mean density  contrast 
inside $r$ is   
\begin{equation}
\nu = \frac{\rho(r)}{\rho_c} = 200 \times \Delta = 200 \hspace{0.1cm} \frac{M(r)}{ M(r_{\rm vir}) } \hspace{0.1cm} \frac{r^3_{\rm vir}}{r^3}.
\label{eq:nu}
\end{equation}

In a universe 
with  matter and vacuum energy densities such that
$\Omega_m (z)  + \Omega_V (z)=1$, the virial radius varies with redshift as  
\begin{equation}
\frac{r_{vir(z)}}{r_{vir(0)}} = 
\frac{  (M (r_{\rm vir}, z)/M_0)^{1/3} }              
{  (\Omega_m  (1+z)^3 + \Omega_V)^{1/3}  }, 
\label{eq:viral}
\end{equation}
where $M_0 = M (r_{\rm vir}, z = 0)$.
 \pcite{wechs02} suggested the empirical relation whereby
\begin{equation}
M_{\rm vir} (z) =   M_0  e^{-  \alpha z}.
\label{eq:wechs}
\end{equation}
This defines  a characteristic halo growth time\footnote{The exponential 
form reflects a two phased evolution;  a period of  rapid accretion followed by slower growth. 
Zhao et al. (2003) have confirmed its basic premises for halos growing 
through almost four orders of magnitudes in mass and three in radius (although they 
do not use this particular form to fit the data).} 
\begin{equation}
\tau_s   \sim \left(\frac{d \ln M}{dt }\right)^{-1} = \frac{-1} {\alpha dz/dt}.
\label{eq:ts}
\end{equation}
The parameter $\alpha \sim 1$ varies systematically with halo mass (but appears constrained 
to the range $0.4 \la \alpha \la 1.6$).
In this paper we fix  $\alpha=1$, noting that our results were found to be 
quite insensitive as to the exact choice of $\alpha$.
Wechsler et al. also find that, for NFW type fits to the density profile,  $c \sim (1+z)^{-1}$, 
with an average value of order $10$ at $z=0$. Their results thus 
generally confirmed the prescription (due to Bullock \etal 2001)  
\begin{equation}
c~[{\rm NFW},z] = \frac{9} {1+z},
\label{eq:cnfw}
\end{equation}
which is used here. We assign a minimal value of 
$c = 9/5$ at high $z$ (because structures with $\rho \sim 1/r$ 
in their outer regions are necessarily far from virial equilibrium; 
the simulations of Bullock et al. 2001; Zhao et al. 2003 and Tasitsiomi \etal 2004 
confirm the presence of a minimal $c$). 
The concentration  parameter of an  M99  fit relates to 
to that of an NFW fit through
$c [{\rm M99}] \approx (c[{\rm NFW}]/1.7)^{0.9}$.

For any given $z$ there is a mass dependent spread in $c$; it is  given, in terms of 
the typical mass scale $M_*(z)$ collapsing at redshift $z$, {\em via} 
$c \sim (M_{\rm vir}/M_*)^{-0.13}$.
 Nevertheless, this relation shows that, even for a halo
that is ten orders of magnitude more massive than $M_*$,
$c$ only changes by a factor of ten. From Fig.~1, this leads to a corresponding change 
in the density contrast of about the same factor or less. 
But since the relaxation time  (cf, Eq.~\ref{eq:famr} below)
$\sim N(r) \tau_D (r) \sim \sqrt{\rho (r)}$ inside any fixed radius; it changes by a factor 
of only a few even for such a highly unlikely mass deviation. 
Trials with different 
normalisations in  Eq.~(\ref{eq:cnfw}), and with the prescription 
proposed by (Eq.~10 of) Zhao et al. (2003), have confirmed this. 
In fact it turns out (Section~4.2) that the relaxation rate
is quite a strong function of radius, 
and will be important at small radii for any reasonable set 
of $c$ values during the evolution. 
The results presented in this paper
assume the quoted form for this relation, ignoring the weak mass 
dependence. Remarkably, our general conclusions
should hold for haloes in any mass range

\subsection{Generalisation of the standard formulation}
\label{sec:diff}

The standard approach to relaxation in gravitational systems has developed 
along the foundation set by \pcite{chandra43}.
In analogy with the case of dilute gases, 
it is assumed that perturbations experienced by a test particle arise  
from independent and local two body encounters. Since the mean field along
a particle's trajectory changes on a timescale comparable to its dynamical
time $\tau_D$, while the timescale of an average encounter is $\sim \tau_D / N$,
this assumption is generally satisfied; encounters with duration $\sim \tau_D$ 
being relatively rare. 
As the number of particles in the representation of a self-gravitating system
 is increased, the effect of strong encounters becomes increasingly 
unimportant; in softened systems,
the subject of cosmological simulations, strong encounters are completely suppressed
if the softening length is of the order of the interparticle  
distance. The standard formulation of relaxation theory therefore assumes 
that the effect of discreteness noise can be modelled in the form of 
weak random encounters. 
Numerical tests  (e.g., \pcite{dub93}),
seem  to vindicate these basic premises.\footnote{Note nevertheless that 
 such work generally   focused on energy 
changes along particle trajectories. 
The response of the trajectories themselves  
(and thus the effect on  quantities like angular momentum) 
can exhibit significantly stronger sensitivity to noise
--- even in the simplest case of deflections 
caused by fixed background particles~(\pcite{at01};~\pcite{ez02}).}

The product  of weak, local and independent encounters
is a diffusion process, whereby
a particle's dynamical variables undergo a random walk around 
their unperturbed values. For a spherical system with isotropic velocities that can 
approximated in terms of a Maxwellian, 
the mean diffusion coefficient within radius $r$ 
\begin{equation}
 \< (\Delta v)^2 \>  = \frac{G^2 \rho m \ln \Lambda}{K \sigma}
\label{eq:sph}
\end{equation}
describes
the average {\em rate} at which the square velocities of particles 
deviate from those of  unperturbed trajectories in a corresponding smooth system 
(with $N \rightarrow \infty$).  
Here $\Lambda$ is the ratio of maximum to minimum impact parameters and
$K=1/15.4$ (according to Spitzer \& Hart 1971).
{\em In quasiequilibrium, at any given instant,}
the average velocity dispersion inside radius $r$ can be calculated from 
\begin{equation}
\sigma^2 (r)  =\frac{3}{r} \int_0^r  \bar{v_r^2} dr = \gamma^2 v_c^2 =\gamma^2 G M/r = \gamma^2 G~\frac{4}{3} \pi \rho r^2.  
\end{equation}
The local one dimensional velocity dispersion $\bar{v^2_r}$ can be evaluated 
by solving the Jeans equation (cf. Appendix B). The above relations thus  
define the weak function of radius $\gamma$ that is used in the calculations below. 
This definition should hold because the virialised region of a cosmological halo remains
near equilibrium through most of its evolution ---  major mergers being relatively 
rare, and even when they do occur the system is out of equilibrium for a time
$\tau_D$ generally much smaller than the (relaxation) timescales of interest.

When the system parameters are changing, as during the growth 
of a cosmological halo, the mass inside any given radius will
be time dependent. As a consequence, there will be  corresponding variations in 
$\rho = \frac{M (r, t)}{4/3 \pi r^3}$ and $\gamma$.                            
Nevertheless, if  the timescale between 
encounters giving rise to the relaxation process ($\tau_e \sim \tau_D/N$) is 
much smaller than the timescale for variation of the system parameters $\tau_s$, then 
local (in time) averages are allowed, and the diffusion coefficient becomes
a well defined function of time. 
From~Eq.~(\ref{eq:ts}), and because
\begin{equation}
 |\dot{z}| = ({1+z})~H
\label{eq:hub}
\end{equation}
and
\begin{equation}
\rho_c = \nu \frac{3 H^{2}}{8 \pi G},
\label{eq:rhoc}
\end{equation}
one finds that $t_s \sim  \frac{1}{1+z} \sqrt{\frac{3}{8 \pi G \rho_c}}$. Therefore, unless
$z$ is very large, $\tau_s > \tau_D  \sim 1/\sqrt{\nu G \rho_c}$ (recall that 
$\nu \ge 200$); with the implication that the temporal locality condition ($\tau_e \ll \tau_s$)
is in fact weaker than that for spatial locality ($\tau_e \ll \tau_D$), required 
for the validation of the  diffusion approach.

The  expression for the mean diffusion coefficient, inside radius $r$ and 
at time $t$, can  then  be written as
\begin{equation} 
 \frac{\< (\Delta v)^2 \> (r, t)}{\sigma^2 (r,t)}  = 
\frac{\sqrt{G \rho(r,t)}}{K \gamma^3(r, t) \sqrt{4 \pi/3}} \hspace{0.1in} \frac{m}{M (r,t)} \hspace{0.1in} \ln \Lambda (r, t).
\label{eq:difft}
\end{equation}
The relative  mean square  perturbation to  particle velocities due to encounters 
with other particles during the halo evolution is
\begin{equation} 
\< v_p^2 / \sigma^2 \>  = \int_{t_f}^{t_0} \frac{\< (\Delta v)^2 \> (r, t)}{\sigma^2 (r,t)} dt,
\label{eq:pert}
\end{equation}
where $t_f$ refers to some chosen initial, reference, formation time
(for any given $r$ measured at $z=0$, we will take the redshift  to be the time when  
$r_{\rm vir} (t_f) = r$) and $t_0$ refers to the end of the simulation (assumed to 
correspond to $z=0$). 

Using~(\ref{eq:difft}) and~(\ref{eq:pert}) one can define 
the relaxation time in  a simulated cosmological halo  implicitly:
\begin{equation}
 \< v_p^2 / \sigma^2 \> (t_{\rm relax})  = \int_{t_f}^{t_{\rm relax}} 
\frac{\sqrt{G \rho}}{K \gamma^3 \sqrt{4 \pi/3}} \frac{m}{M} \ln \Lambda dt = 1.
\label{eq:genrelax}
\end{equation}
If $t_{\rm relax} \la t_0$ the dynamics can be expected  
to be completely dominated by discreteness noise. This is in line 
with the standard application to time independent systems, where $t_f = 0$ and
$ \< v_p^2 / \sigma^2 \> $ is assumed to be time independent; the 
the result are the familiar expressions for the relation time 
\begin{equation}
t_{\rm relax}  = K  \frac{\sigma^3}{G^2 \rho m \ln \Lambda} =
\frac{K  \gamma^3 \sqrt{\frac{4}{3} \pi} }{\ln \Lambda \sqrt{G  \rho}} \frac{M}{m}
\hspace{0.06in} \sim~0.1~\frac{N}{\ln \Lambda}~\tau_D
\label{eq:famr}
\end{equation}
(where $N(r)$ and $\tau_D (r)$ are the particle numbers and mean dynamical time 
inside $r$).

To fix the value of $\Lambda$, we assume that the resolution of the simulations
in question corresponds to the local interparticle distance. The average of this
quantity within radius $r$ will define the minimum impact parameter
$b_{\rm min} = (\frac{4}{3} \pi r^3/N(r))^{1/3}$.  The maximum impact parameter
will be taken to correspond to the  virial radius. We therefore have
\begin{equation}
\Lambda (r,t)  = \frac{b_{max}}{b_{min}} = 
\left(\frac{3}{4 \pi}\right)^{1/3} N^{1/3}(r,t) \hspace{0.1in} \frac{r_{vir}(t)}{r}.
\end{equation}

\section{Simple estimates}
\label{sec:simple}

In the next section we will numerically evaluate the integral in Eq.~(\ref{eq:genrelax}), 
invoking the time dependence appropriate in 
the currently favoured  $\Lambda$CDM cosmology. Some insight can however be gained by 
defining  characteristic relaxation times for an evolving cosmological halo and 
for its substructure. This is the subject of this section.

\subsection{A characteristic relaxation time for cosmological haloes}
\label{sec:simple_main}

\begin{figure*}
\psfig{file=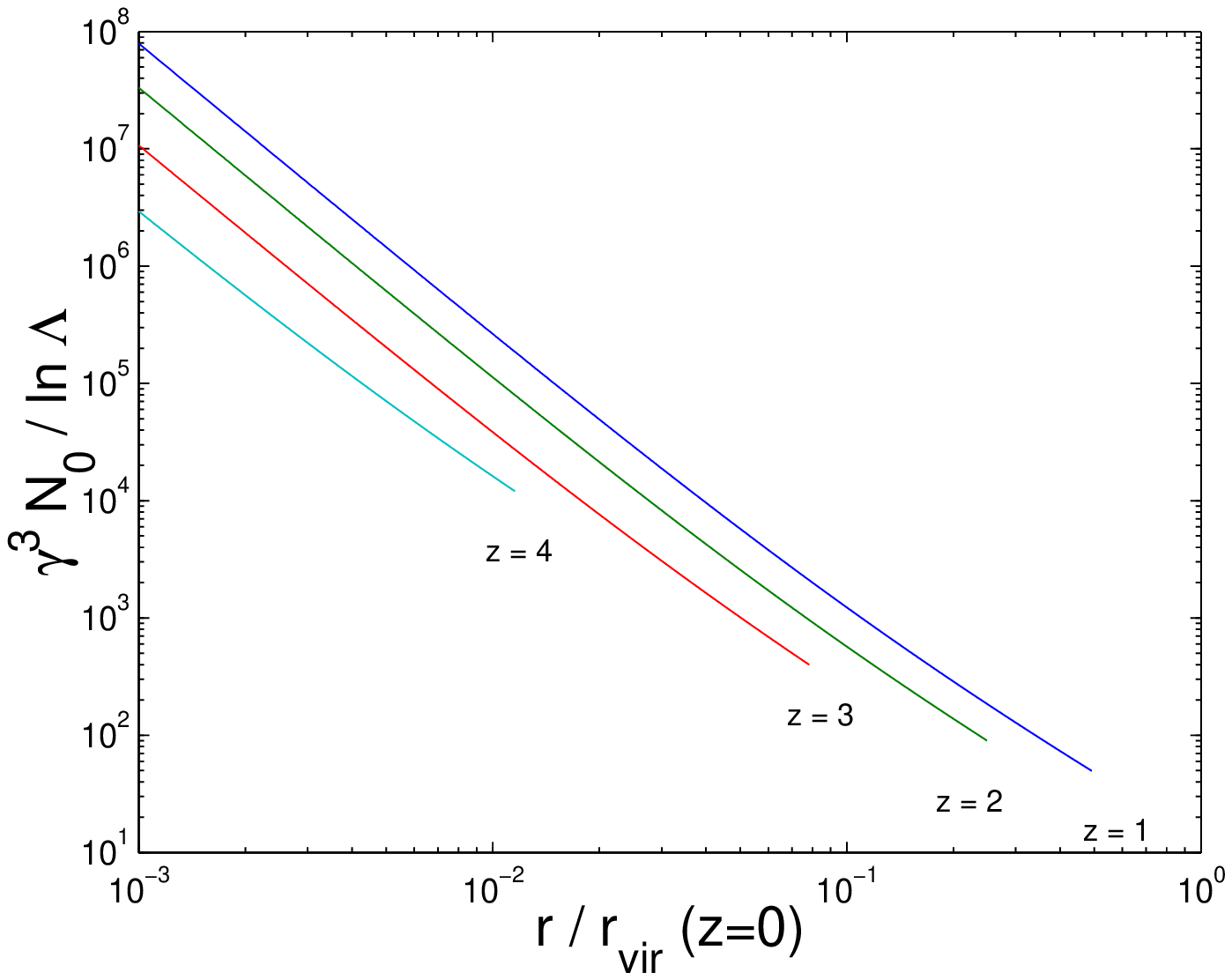,width=87mm}
\psfig{file=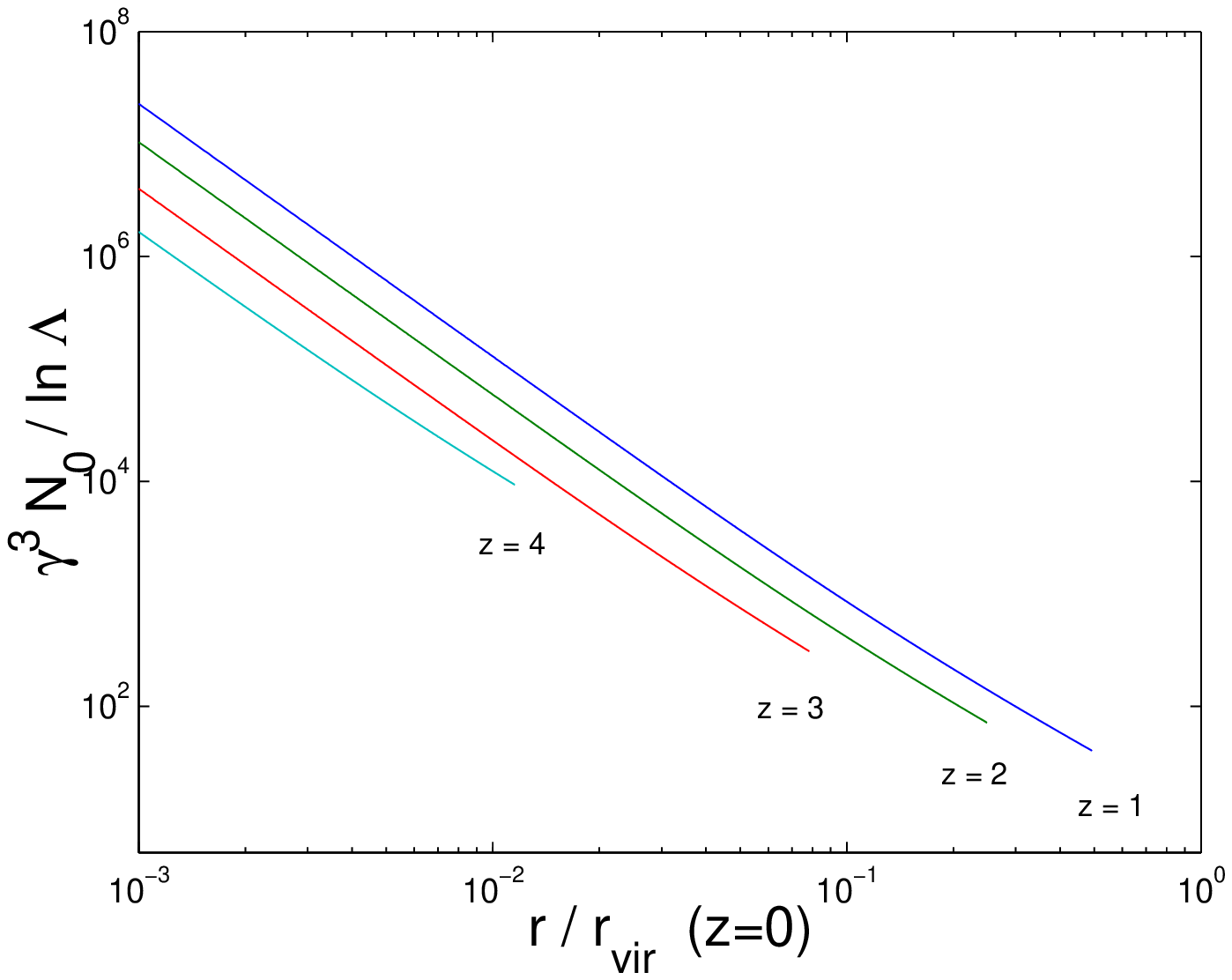,width=87mm}
%\onecolumn
\caption{Number of particles $N_0$ required to compose a final main halo at $z=0$ 
so that, at indicated redshifts, the mean relaxation time 
will be larger than the characteristic time for the 
the halo to increase its mass at that redshift (see Eq.~\ref{eq:ts}). 
Radii are rescaled to units of the virial radius at $z=0$. The plot
on the left corresponds to NFW haloes while that on the right to the M99 profile.
Note that $\gamma^3 / \ln \Lambda \sim 1$}
\label{fig:doublemass}
\end{figure*}

We consider first the evolution of the 
parent halo, or most massive progenitor, and assume that all mass is added to it
in the form of a smooth component. Of course, in reality, accretion of material 
takes the form of clumpy subhaloes, but most of their mass is rapidly stripped, so that,
at any given time, there is usually only $10-20 \%$ of the mass of the halo in 
the substructure. Their internal relaxation is dealt with separately below.

If  the relaxation time within
some radius $r$  is smaller than  the timescale $\tau_s$ 
for the mass distribution to significantly evolve,  it is
likely that the system is heavilly affected by discreteness noise within radius $r$
--- the rationale being that if the local
relaxation time is small compared to the time it takes for more particles to be 
added to the system (thus decreasing the relaxation rate) then discreteness 
noise will have a significant effect. 

A  local relaxation time can be obtained by freezing the 
system at some time $t = t(z)$ and using~(\ref{eq:famr}).
 Comparing the two timescales one can write
\begin{equation}
R= t_{\rm relax}/\tau_{\rm s} =
 \frac{K \alpha \gamma^3 \sqrt{\frac{4}{3} \pi} }{\ln \Lambda} \frac{|dz/dt|}{\sqrt{G  \rho}} \frac{M}{m},
\end{equation} 
and eliminate $dz/dt$ by invoking~(\ref{eq:nu}),~(\ref{eq:hub}) and (\ref{eq:rhoc}) to 
get 
\begin{equation}
R (r,z)= \frac{0.087 \sqrt{2/\nu} \pi \alpha \gamma^3 }{\ln \Lambda}  (1 + z)  N (r, z),
\end{equation}
where (by Eq.~\ref{eq:wechs}) the number of particles inside $r$ at redshift $z$
can be expressed in terms of the final number of particles with which the halo is 
resolved $N_0 = M_0/m$:
\begin{equation}
N(r,z) = \frac{M}{ M_0} \hspace{0.1in} N_0~e^{- \alpha z}.
\end{equation}
The mass ratio entering into this last relation can be calculated
{\em via} either  Eq.~(\ref{eq:massnfw})  or Eq.~(\ref{eq:massmoore}).

 The dimensionless relaxation time  $R$ has to be significantly 
greater than unity, at all $z$, if the simulated halo can be considered  
free of artificial relaxation inside radius $r$.\footnote{Note that, because 
the  virial radius and concentration are time dependent, 
accreted mass is not deposited  uniformly; there is a 
preference, especially at later times, for mass increase in the outer regions (e.g., Zhao et al. 2003).
 However this implies 
that the characteristic time for mass change in the inner regions, where most of 
the relaxation occurs, is larger than that predicted by Eq.~(\ref{eq:ts}). 
The condition $R > 1$ thus constitutes a minimal requirement.}
What is required therefore is that 
\begin{equation}
N_0  > 11.5 \hspace{0.06in}
 \frac{\sqrt{\nu/2}}{\pi \alpha \gamma^3} \hspace{0.06in} \ln \Lambda   
 \hspace{0.06in} \frac{M_0}{M} \hspace{0.06in} \frac{e^{\alpha z}}{1+z}.
\label{eq:num}
\end{equation}
In  Fig.~\ref{fig:doublemass} we show the variation of the of the quantity 
$(\frac{\gamma^3}{\ln \Lambda}) N_0$ as a function of 
radius at different redshifts.  The radii are expressed in terms of the 
final virial radius (at $z=0$), by transforming them using Eq.~(\ref{eq:viral}).
The cutoffs in the 
curves correspond to the virial radius at the denoted redshift (that is the 
maximum size of the halo considered at that redshift).

From this figure it is apparent that, provided   $\gamma^3 \sim \ln \Lambda$, 
that artificial relaxation may have a dominant effect inside the inner 
$\sim 1 \%$ of the final virial radius for $N_0 \sim 10^6$.  
Note that the demand for larger particle number is most stringent at larger 
times (smaller $z$). This is because the characteristic growth time $\tau_s$ is a steeply 
increasing function of $z$.
The figure also suggests that, for fixed $\gamma$,
 the relaxation effects are more pronounced in the inner regions of  NFW 
haloes, compared  to the M99 case.  
Nevertheless,  as we will see in Section~\ref{sec:relaxpar}, the introduction of variation in $\gamma$
reverses this situation, since $\gamma$ is significantly 
larger in the inner regions of NFW systems (this follows from the variations
of the ratio of velocity dispersion to circular speed shown in Fig.~B1).

\subsection{Characteristic relaxation time for substructure}
\label{sec:simple_substruc}
\begin{figure}
\psfig{file=  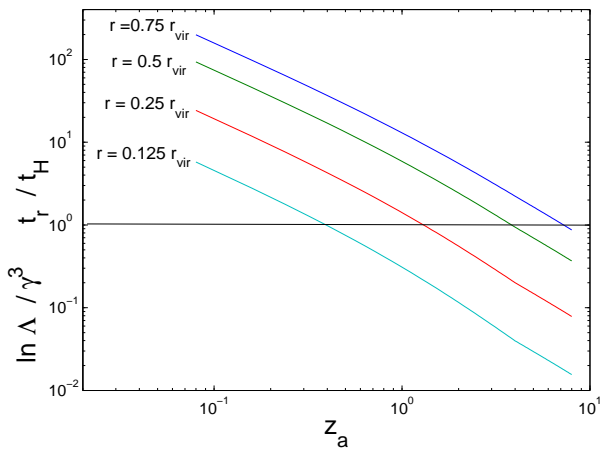,width=87mm}
%\onecolumn
\caption{Relaxation time of substructure haloes consisting of a thousand particles
accreted at redshift $z_a$.
The timescales are expressed in terms of the time $t_H$ elapsed between epochs 
$z_a$ and $z=0$ and are given at different fractions of the final virial radius.
Values smaller than unity correspond to situations where the dynamics is completely
dominated by artificial relaxation. Haloes are assumed to be of the NFW type} 
\label{fig:simple_substruc}
\end{figure}

We consider a subhalo accreted at redshift $z_a$ and remaining
a separate dynamical system  up to $z=0$ --- i.e., it stops growing,  its outer
regions are in fact stripped, but it keeps a dynamically distinct core.
For this purpose  we  exploit the possibility of relating the relaxation and the local 
Hubble time; and  in order to get simple closed form estimates, 
we will neglect relaxation at times $t < t (z_a)$ and focus on relaxation effects 
since accretion.
From equations~(\ref{eq:famr}), (\ref{eq:hub}) and (\ref{eq:rhoc}) we can define 
a characteristic relaxation time
\begin{equation}
t_{\rm relax} (r,z_a) = \frac{0.087 \pi \gamma^3 \sqrt{2/\nu}} {\ln \Lambda} 
\hspace{0.1 in} \frac{M}{m} \hspace{0.1in} H^{-1} (z_a),
\end{equation}
where $M,~\nu,~\gamma$ and $\Lambda$ are all measured inside radius $r$ at $z = z_a$.  
This  timescale should be compared to the time interval separating the accretion 
epoch $t (z_a)$ and the end of the simulation at $t_0 = t (0)$. This 
 determines a characteristic substructure relaxation time given by 
\begin{equation}
t_{rsub}(r, z_a) = \frac{t_{\rm relax} (r, z_a)}{t_0 - t (z_a)}.
\end{equation}

Now further assume 
that $t (z) = \frac{2}{3} H$ and  $z = (t_0/t(z))^{2/3} -1$, as is appropriate 
for an Einstein de-Sitter universe (in the next section we verify our results 
by undertaking the full calculation, including pre-accretion evolution,
 in the currently favoured ``concordance'' 
cosmology\footnote{Note that while  $t = t(z)$  for $\Lambda$CDM
is quite different from that in an Einstein-de-sitter model
the {\em ratios} $t(z_1)/t(z_2)$, entering into the 
equation below, differ by a factor of at most by 
$\sim 3 \%$ for  $10  \ga z \ga 1$ and by $\sim 30 \%$ up to $z=0$.}). 
In this case  we have
% 
%\begin{equation}
%t_{\rm relax} = \frac{3}{2} \hspace{0.1in} \frac{0.087 \pi \gamma^3 
%\sqrt{2/\nu}} {\ln \Lambda} \hspace{0.1in} \frac{M}{m} 
%\hspace{0.1in} t (z_a).
%\end{equation}
%
%
\begin{equation}
t_{rsub}(r, z_a) = \frac{3}{2}   \hspace{0.1in}
 \frac{0.087 \pi \gamma^3 \sqrt{2/\nu}} {\ln \Lambda}  \hspace{0.1in} 
\frac{N(r, z_a)}
{(z_a + 1)^{3/2} -1}.
\end{equation} 
If, within some given radius, this quantity is less than unity the implication 
is that, by $z=0$, the dynamics has become dominated by artificial discreteness 
noise.

In Fig.~\ref{fig:simple_substruc} we reproduce several plots where this dimensionless 
relaxation time, thus defined, 
is shown for haloes assumed to have $N (r_{\rm vir}, z_a) =1000$ when they are  
subsumed into
their parent structures at different $z_a$ (in the case of satellites haloes we only 
reproduce the results for NFW haloes, those for the  M99 haloes are very similar).
 As can be seen, especially at relatively small
radii, the relaxation time  can be significantly smaller than the subhalo lifetime.
Furthermore, it is to be noted that it will be this inner core of the subhalo that will
survive stripping (see also Fig.~\ref{fig:satnfw}  for more detail concerning 
the radial distribution of the relaxation time and Section~\ref{sec:consub} for a  
discussion of some possible consequences).

\section{Direct Calculations}
\label{sec:full}
\begin{figure*}
\psfig{file = 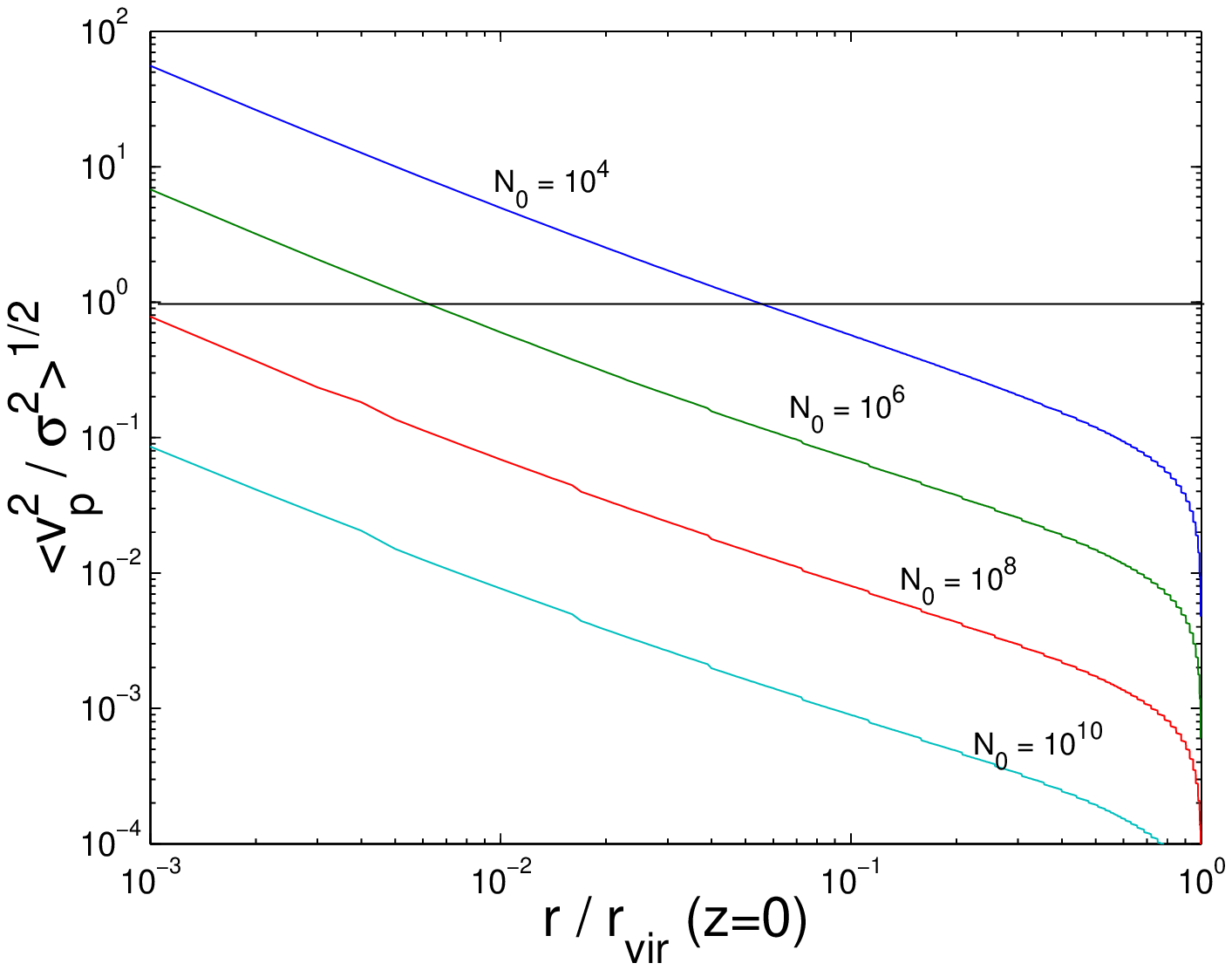,width=87mm}
\psfig{file = 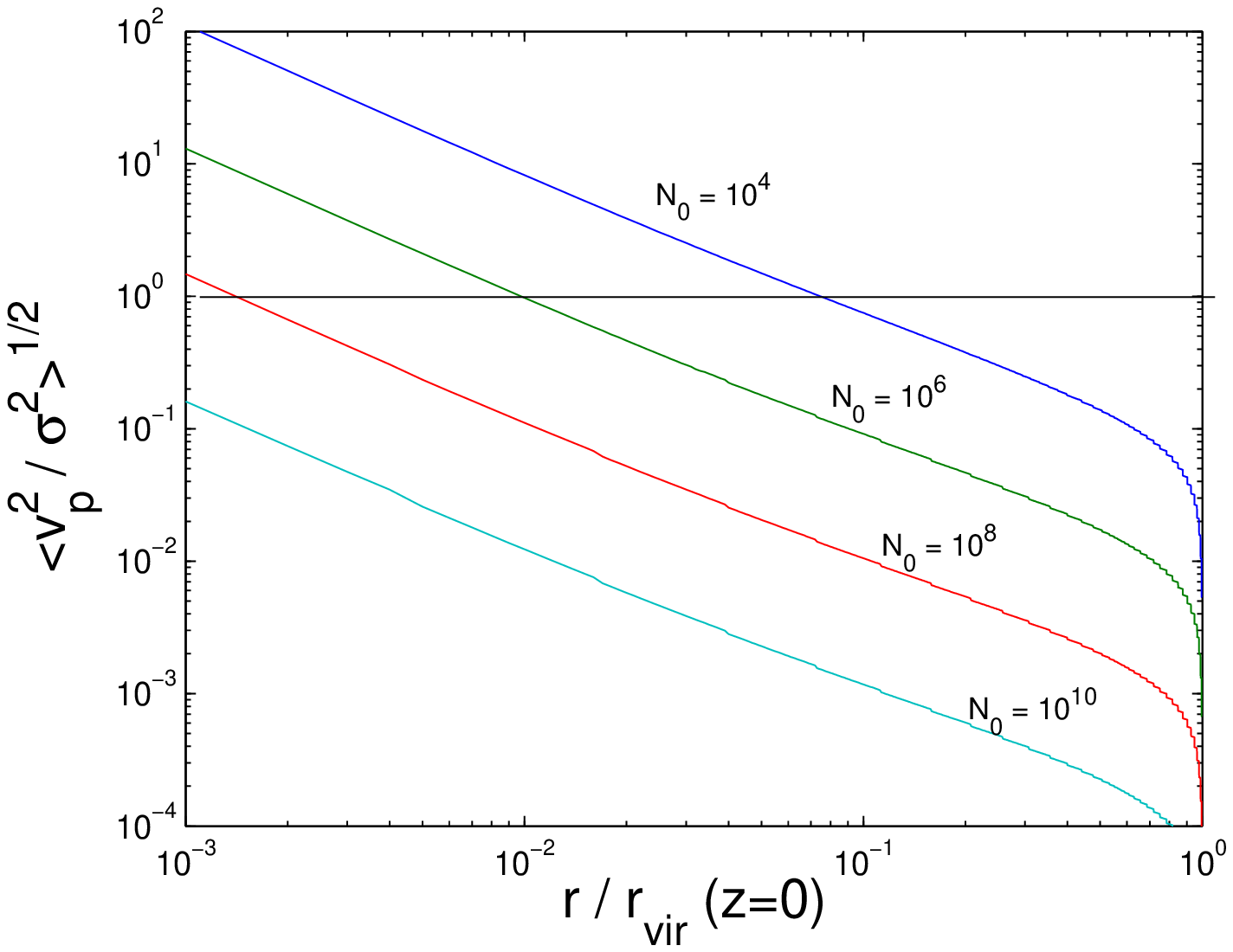,width=87mm}
%\twocolumn
\caption{Expected RMS perturbation to the velocity resulting from particle noise in 
evolving most massive progenitor cosmological haloes. It is  
evaluated at a given fraction of the final virial radius
by integrating Eq.~(\ref{eq:msdirect}), from the time
when that fraction was  equal to the  whole  virial radius at epoch $t= t_f$ up to $z=0$
($t=t_0$). Results for NFW haloes are shown on the left while those corresponding 
M99 ones are on the right}
\label{fig:relax}
\end{figure*}
\subsection{General considerations}
From equations~(\ref{eq:difft}) and~(\ref{eq:pert}), the expected relative mean square perturbation 
due to discreetness imposed on to a test particle moving  within an 
evolving cosmological halo is
\begin{equation}
\<v_p^2 / \sigma^2 \> = \frac{1}{2 K} \sqrt{\frac{3G}{\pi}}
\int_{t_f}^{t_0}  \ln \Lambda(r,t) \hspace{0.025in} \frac{\sqrt{\nu (r,t)}}{\gamma^3 (r,t)} \hspace{0.025in} \frac{\sqrt{\rho_c (t)}}{N (r, t)}\hspace{0.025in} dt.
\label{eq:msdirect}
\end{equation}

The integral is evaluated within a given fixed 
fraction of the {\em final} virial radius $r_{vir}(t_0)$ --- with the consequence that the radial
coordinate contains an implicit time dependence (i.e, $r = r(t)$).
The relations in Section~\ref{sec:massmod} can be used to 
determine this  dependence, as well as that of the other quantities entering 
into~(\ref{eq:msdirect}),
provided the evolution of redshift  as a function of time  is known.
For a flat universe with cosmological 
constant, the required transformations are  given in Appendix~A
(in the calculations below we use $\Omega_V = 0.7$ and $h=0.7$).
   Since the virial radius increases with time, a given fraction of the virial radius at 
$t = t_0$ will correspond to the whole virial radius at some earlier time. This naturally
determines the lower limit of integration, the  formation time  $t_f$ --- for some fixed fraction of 
$r_{\rm vir} (z = 0)$, it corresponds to a redshift $z_f$ where this 
fraction is 
equal to whole virial radius $r_{\rm vir} (z_f)$.

\subsection{Relaxation of the most massive progenitors}
\label{sec:relaxpar}

In Fig.~\ref{fig:relax} we show  the RMS perturbation due to particle noise, 
as a function of radius, expected for a halo with mass evolving according to 
Eq.~\ref{eq:wechs} (with $\alpha = 1$) for several final values of the final 
total particle number $N_0$.
They are obtained by solving~(\ref{eq:msdirect}) using an
adaptive integrator (NAG D01AJF) with a tolerance of $10^{-4}$.

 An interesting characteristic of the plots in Fig.~\ref{fig:relax} 
is the perfect power law
behaviour of $\< v_p^2 / \sigma^2 \>^{1/2}$ over most of the radial interval.
This is a consequence of a curious property of cosmological haloes, already noted 
by~\pcite{tanav01}; namely the power law form of the  phase space density. It is this phase space 
density, $\sim \rho/\sigma^3$, that determines the rate of relaxation. 
This property  enables one to deduce a particularity simple fit for the variation of the 
relative RMS perturbation:
\begin{equation}
\< v_p^2/\sigma^2 \>^{1/2} \approx~10^{-3}~\sqrt{\frac{10^8}{N_0}}~\frac{r_{\rm vir}}{r}, 
\label{eq:fitp}
\end{equation}
where $r/r_{\rm vir}$ refers to the fraction of the virial radius at $z=0$.
The number of relaxation times inside radius $r$ can be counted as follows:
\begin{equation}
n~(t_{\rm relax}) = \< v_p^2/\sigma^2 \> \approx 10^{-6}~\frac{10^8}{N_0}~\left(\frac{r_{\rm vir}}{r}\right)^2,
\label{eq:fitr}
\end{equation}
which of course needs to be much smaller than one if artificial relaxation is to be negligible within 
radius $r$.

For  $N_0 \la 10^6$ then,  particle motion is  expected to be 
entirely dominated by noise in the inner percent of  the virial radius;
regions bounded by radii an order of magnitude smaller still having undergone 
$\sim 100$ relaxation times. 
These results are  in agreement with the simple estimates presented in 
Section~\ref{sec:simple_main} (with the difference that the effect on M99 haloes are
larger here, because  variations in $\gamma$ is taken into account).

The inner $0.1 \%$ is one relaxation time old even for  $N_0 = 10^8$. Indeed, a final 
particle number inside the halo's virial radius of $N_0 = 10^{12}$ is required for 
noise to be reduced to the level of  a few percent there. 
For a million particles this noise level is only reached at almost a third of 
the final virial radius; and  this is larger than the average 
halo scale length $r_s = r_{\rm vir}/c$ (c.f. Eq.~\ref{eq:cnfw}).

\begin{figure*}
\psfig{file = 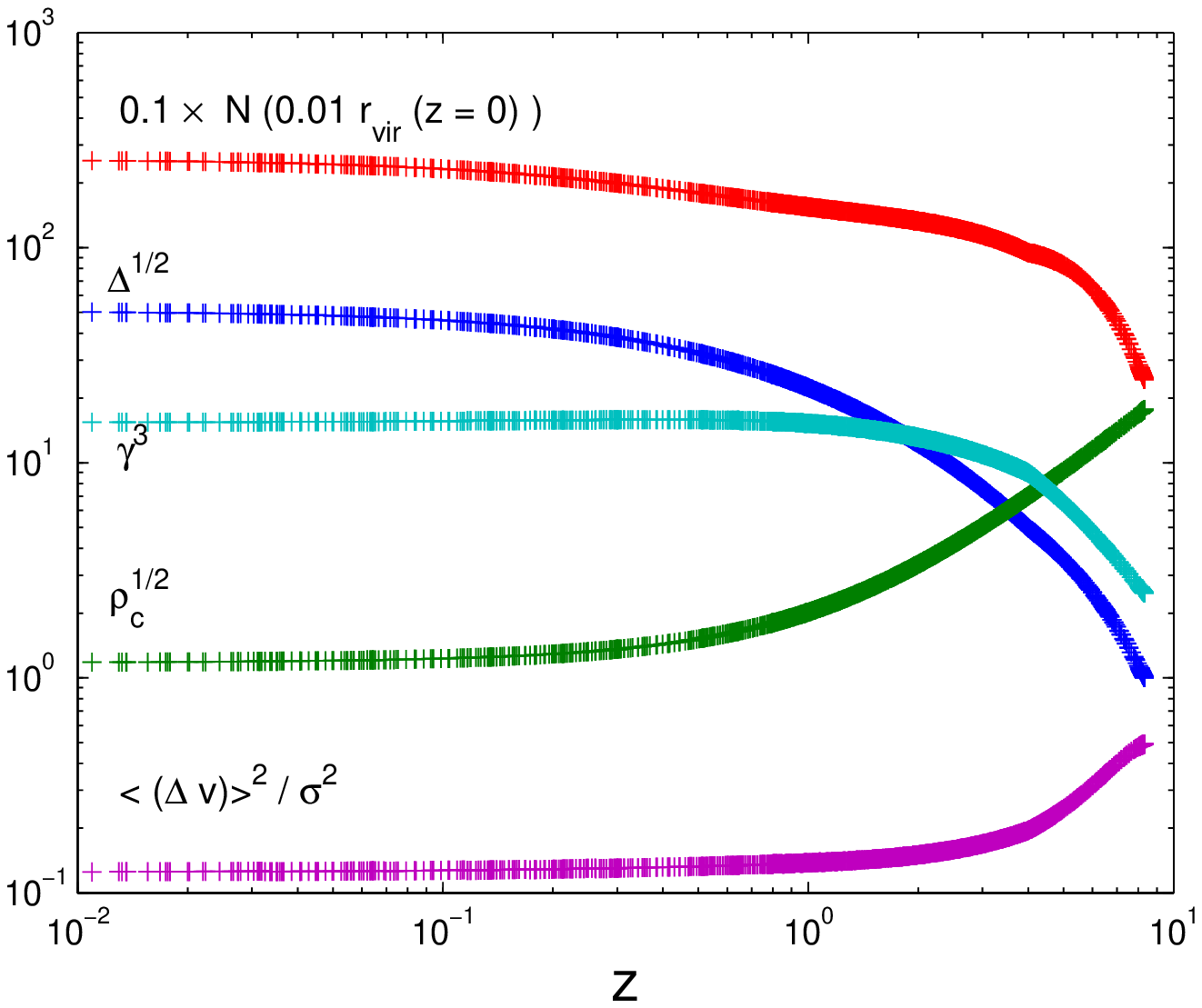,width=87mm}
\psfig{file = 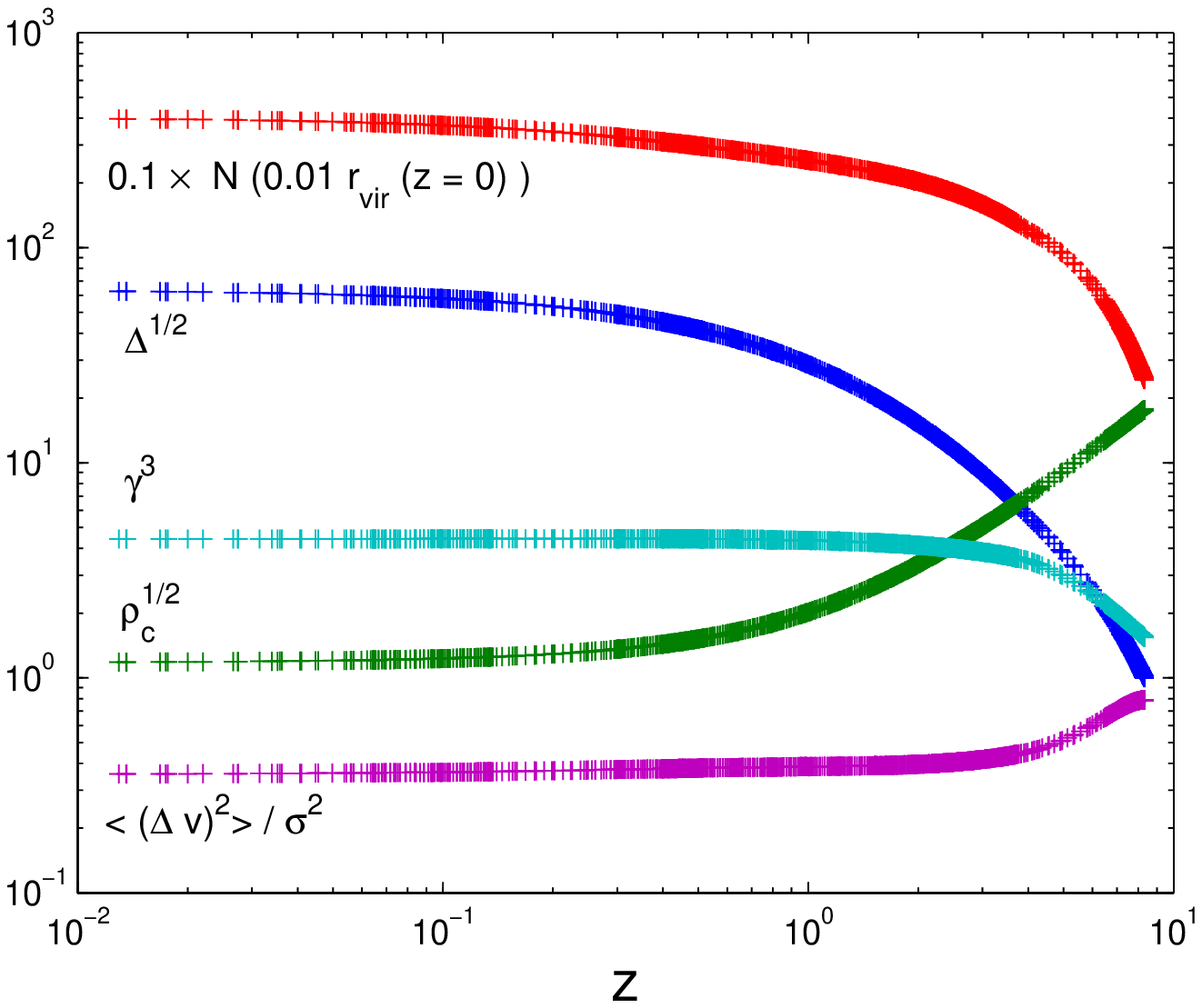,width=87mm}
%\onecolumn
\caption{Instantaneous values of the perturbation to the velocities that arises
from discreteness noise  inside $1 \%$ of the final virial radius, 
and the corresponding magnitude of terms figuring in Eq.~({eq:msdirect}) that 
contribute to this quantity for  NFW haloes (left) and M99 profiles. 
All contributions
are of similar magnitude for the two profiles, 
except for the ratio of velocity dispersion to circular velocity, which 
is larger at small radii (hence lower redshift) in the NFW case. This leads
 to a somewhat smaller perturbation}
\label{fig:contributions}
\end{figure*}

 Fig.~\ref{fig:contributions} shows, for the inner percent of the final virial radius, 
the $z$-evolution of the relaxation rate (the integrand in \ref{eq:msdirect})
along with the principal components determining it. 
 It shows  that the relaxation rate 
decreases with redshift, for both the NFW and M99 profiles --- the effect 
 being more  pronounced in the 
former case because the ratio of velocity dispersion to circular velocities that 
determines $\gamma$ decreases faster with radius (Appendix~B); and  at smaller $z$
because, for a given fixed fraction of the 
final  virial radius, smaller radii are probed (recall that at $z_f$ the whole
virial radius  corresponds to the inner one percent at $z=0$).

Finally, note that, since new particles  are continuously 
being introduced into the region,  the way in which 
particles are affected by relaxation will differ, with the probable consequence 
that relaxation driven evolution can be expected to be  different   
than in a static NFW type system.

\subsection{Relaxation of subhaloes}
\label{sec:relaxsub}
\begin{figure}
\psfig{file=  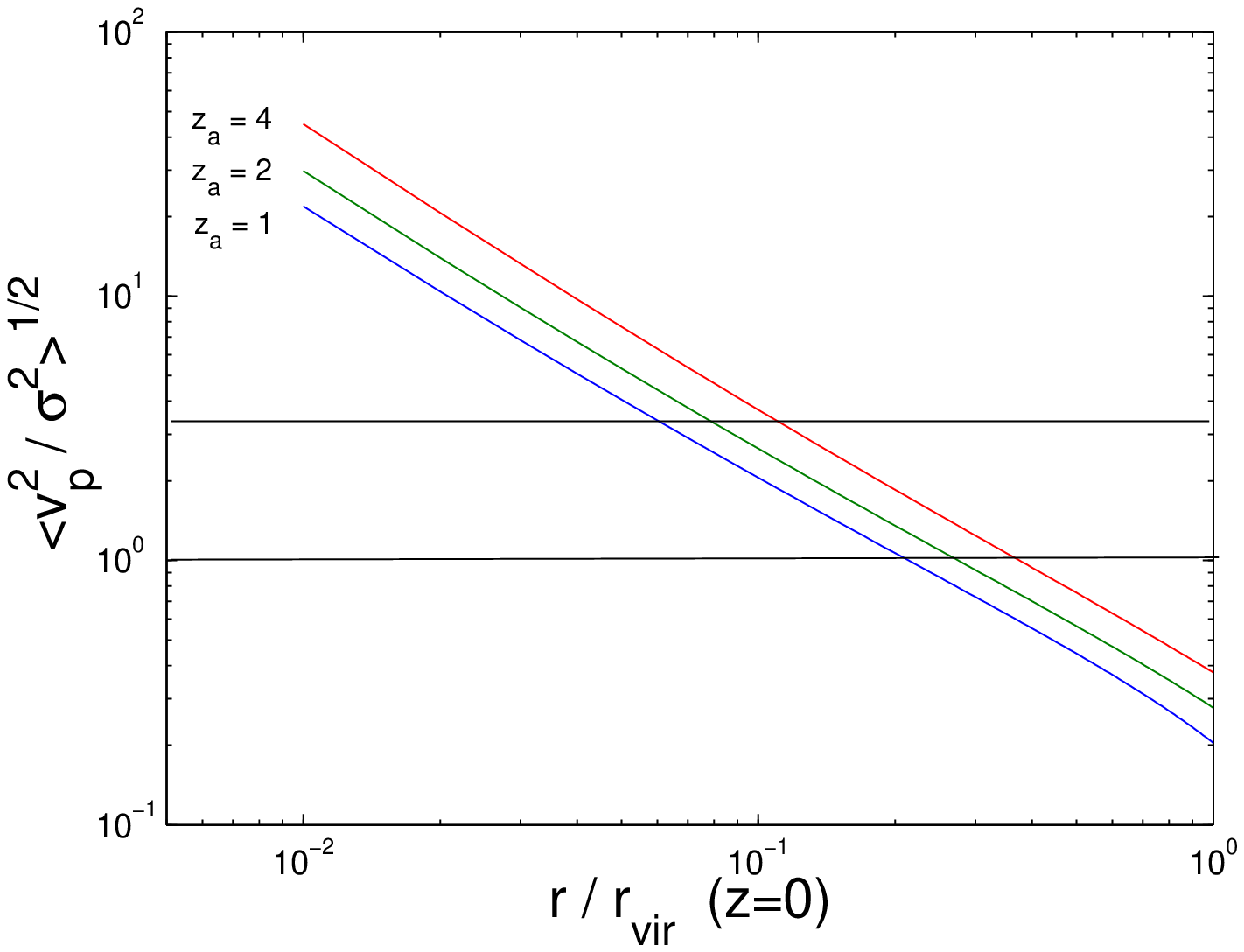,width=87mm}
%\onecolumn
\caption{RMS perturbations to the velocity of satellite haloes composed 
of a thousand particles when they stop growing due to incorporation into 
a larger structure at the redshifts shown (up to which they are assumed to 
grow in mass, and hence particle number, according to Eq.~\ref{eq:wechs}).
Results are obtained using Eq.~(\ref{eq:msdirect}) and correspond to subhaloes with 
NFW profiles. Horizontal lines show perturbations corresponding to (from bottom up) 
one and ten relaxation times}
\label{fig:satnfw}
\end{figure}

In Fig.~\ref{fig:satnfw}  we show the RMS perturbation in velocities expected as a result of 
particle noise for a subhalo composed of a thousand particles when it stops growing 
(after  being incorporated into a larger structure). Before this, we 
assume that its mass grows at the exponential rate described by Eq.~(\ref{eq:wechs})
with (as always)  $\alpha = 1$.
These results are in close agreement with the simple estimates derived in 
Section~\ref{sec:simple_substruc}; they reinforce the conclusion 
that relaxation can be a significant factor determining the structure of 
subhaloes in simulations, their distribution and their effect on the parent
halo structure (Section~\ref{sec:consub}).

\section{Conclusion}
\label{sec:conc}

This  paper  presented  an attempt at assessing the importance of particle noise in 
the evolution of collapsed cold dark matter structures, by generalising the diffusion formulation
first proposed by Chandrasekhar to the case when the mass distribution and radial extent of 
the system under consideration are variable. Although a most massive progenitor of a $z=0$ 
halo will typically have increased its collapse mass by more than an an order of magnitude  by 
accreting  subhaloes, at any given instant $\ga 80 \%$ of its mass is 
in a smooth component, most material in subhaloes having been stripped (e.g., Gao et al. 2004).  
This enables one  to divide the  analysis in two parts. In this context, 
`parent' haloes, i.e ones characterised by continuously increasing (virial) mass and 
radius throughout their evolution, were grown according 
to the empirical formula  discovered by Wechsler et al. (2002).
Subhaloes were treated differently, by assuming that their growth is arrested at accretion.
Stripping was not explicitly taken into account. But since it is in the inner region of subhaloes 
that relaxation is most significant, this does not appear to represent a major idealisation --- as 
high resolution  simulations, specially designed for the purpose of 
examining the issue, show  the inner regions of stripped haloes  to be largely unaffected 
(Kazantzidis et al. 2004).

We have derived  simple closed form formulas for the characteristic 
relaxation times (Section~\ref{sec:simple}), and also integrated the relevant equation 
characterising the expected RMS perturbation to the velocity directly (Section~\ref{sec:full}),
The predictions of the two approaches agree quite closely. 
In what follows we summarise our results and sketch some 
possible consequences.

\subsection{Relaxation in parent haloes}

Our results were not found to be significantly affected 
by the exact form chosen for the evolution of the concentration parameter $c$ or the parameter $\alpha$
characterising the mass growth rate (cf. Eq.~\ref{eq:wechs}), both quantities that are 
dependent on the final halo mass (in physical units), so this latter quantity does not enter
directly into our presentation here, which is concerned only with the relaxation of particles 
{\em after} they have been accreted onto the most massive progenitor and are subsequently
part of its smooth component (in the next two subsection we consider relaxation inside 
substructure, and briefly comment on the expected mass dependence of noise-induced relaxation in terms 
of merger history).

The power law (with index $\sim -2$) radial variation of 
the phase space density of cosmological haloes ensures a 
steep dependence of the relaxation rate, a linear function of this quantity, 
on radius. The perturbation due to discreetness noise is adequately 
quantified by the empirical fits given in Section~\ref{sec:relaxpar}.

A test particle that is present inside the inner percent of the final virial radius,
from the point in time when the virial radius of the halo  was equal to this   
fraction,  will experience a relative  RMS perturbation to its  
velocity  $\sim 1$  when the final halo is resolved with $10^6$ particles --- 
implying that its dynamics is completely corrupted 
by discreteness noise. In the very inner few thousandth of the 
final radius the RMS perturbation due to discreteness noise is an order of magnitude 
larger, meaning that these regions are of the order of a hundred relaxation times old.

In an evolving cosmological halo, particles are continually added within any given radius;
it is also likely that some can gain energy (e.g. by interaction with sinking 
satellites; cf. El-Zant et al. 2004, El-Zant 2005) and move into trajectories with larger average radii. 
 Particles that are accreted at a later stage  are more mildly affected by relaxation.
At any given fraction of the final radius therefore, there will be a range in the 
severity of artificial relaxation, depending on the accretion epoch of individual particles.
This somewhat complex situation implies that effects of relaxation may be rather different 
than the familiar situation of fixed-mass systems, and therefore difficult to detect.

Our calculations  thus represent a   
quantification of the magnitude of the perturbation of the discreteness noise,  
rather than its effect on the detailed dynamics and macroscopic structure.  
While  it  is quite probable  that the significant role for artificial
relaxation predicted by our analysis is reflected in the contradictory claims 
concerning the convergence and shape of the inner density profile of CDM haloes
(Navarro et al. 2004; Diemand et al. 2005), with the  predicted effect for static systems 
being a weakening of the cusp, further work is needed to determine the consequence in the 
case of evolving cosmological haloes.

The RMS perturbation to the velocities decreases as $\sim 1/\sqrt{N}$, 
implying that a parent halo needs to be resolved with $N \ga 10^{10}$ particles at $z=0$ 
if discreteness noise is not to dominate the dynamics of the inner $0.1 \%$ of the virial 
radius; and if it is to be negligible ($\<v_p^2/\sigma^2\>^{1/2} \la 1 \%$) in the whole  region
where  $r \ga 0.1 r_s \sim 0.01 r_{\rm vir}$. 
While perturbations of order unity are likely
required for significant energy relaxation and restructuring of the azimuthally 
averaged density profile, much smaller perturbations may  affect 
particle trajectories,  with accompanying
modification (isotropisation) of the velocity distribution and loss of
triaxiality in spatially asymmetric systems (e.g., \pcite{meval96}; see also Section~\ref{sec:diff}).\footnote{The reason 
this happens is due  to a transition from regular to chaotic motion, which is very sensitive 
to perturbation. A very simple example of this phenomenon is the case of a pendulum on a  
`separatrix' trajectory passing near the unstable (upper) equilibrium. Small 
perturbations can turn (regular) trajectories rotating in one direction into ones rotating into
the opposite direction, or oscillating without a definite sense of rotation, or even
transit between all these different possibilities (chaotic trajectories).}

\subsection{The relaxation of substructure}
\label{sec:consub}

The results presented in this paper  suggest that the 
situation with substructure is still more severe. A  parent halo that is identified with 
$N_0 \sim 10^7$  has (from Eq.~\ref{eq:wechs})
$\sim 10^6$ particles at $z \sim 2$. Suppose  a subhalo with  $N_{\rm sub} = 0.001 N$ is then accreted.
Assume that between redshifts two and one this  $1000$-particle halo will be stripped of 
$\sim 90 \%$ of its mass (as would be expected from Gao et al. 2004 Fig.~14). For an 
NFW halo this also corresponds to a truncation of about $90 \%$ in radius. But according to
Fig.~\ref{fig:simple_substruc}, this inner region will be significantly affected by relaxation. 
In fact, by $z=0$, it is expected to have undergone enough relaxation times to have approached 
core collapse! But this fate is only likely if it is not, in the meantime, dissolved by 
stripping.\footnote{Note that, because  relaxation  is quite a  sensitive function of radius ($t_{\rm relax} \sim r^2$
from Fig.~\ref{fig:satnfw} and Eq.~\ref{eq:fitr}) this same argument  can be made for more central 
regions of more massive subhaloes (e.g., for the inner $1 \%$ of a subhalo consisting of $10 \%$
the mass of mass of the parent halo, accreted at $z \sim 2$ when the parent halo had $N \sim 10^6$), 
or more extended regions of less massive ones.}

What effect is the significant relaxation expected to have on stripping? Cosmological haloes have a 
`temperature inversion' in their inner velocity distribution (cf.~Appendix~B). Relaxation therefore causes initial 
expansion of the inner region, forming an  isothermal core, before it recontracts like a relaxing globular
cluster (e.g.,~\pcite{hayashetal03}). In the first phase, the core becomes less dense and more easily stripped;
in the second the situation is reversed. Stripping is most efficient for subhaloes that venture near the centre
of the parent, experiencing strong tidal forces. If, while the outer regions of a subhalo are stripped during
successive passages, the inner region relaxes to a more diffuse density distribution, further stripping
will be accelerated and the core may completely dissolve. Conversely, if the stripping is slow, there may be 
sufficient time for core contraction to take place. Further stripping is subdued and core dissolution becomes
less likely.

If, as is suggested in several studies (e.g., \pcite{sywt98,dekev03,dekar03}), the structure of  the halo 
profiles found in numerical simulations is dependent on the  interaction between the subhaloes and their parents, then 
it is clear that  such significant re-engineering of the subhaloes can be of major importance in determining the parent halo
profile. It obviously also has implications for the number  and spatial and orbital distribution of  substructure.
Given the discussion above, one would expect artificial relaxation to enhance the number of haloes that do not 
venture into the inner regions (i.e.,  those accreted on less  eccentric orbits) and decrease the number 
of those that do venture there.

\subsection{The N-scaling of the substructure-dominated relaxation time}

  The expectation from the results just outlined is that, beyond the inner
percent or so of the final  virial radius, relaxation will be 
dominated by particles inside substructure, or which have spent considerable 
time inside subhaloes before being stripped. 
 
Simulations suggest that substructure has a mass function such that  
$dn/dm \sim m^{-9/5}$, quite independently of redshift (e.g., Gao et al 2004).
If one ignores the logarithmic dependence, the relaxation time when expressed in units of dynamical time is, 
for any given subhalo,  a linear function of the number of particles in it.
Consider then  a mean relaxation time  averaged 
over subhaloes and  normalised over an averaged dynamical time,
\begin{equation}
\< t_{ra} \> \sim   \frac{N_0 \int_{m_{\rm min}}^{m_{\rm max}} m \times m^{-9/5} dm}{\int_{m_{\rm min}}^{m_{\rm max}}  m^{-9/5} dm},
\end{equation}
where  
the integration limits correspond to the least and most massive subhalo present.  When the latter
is orders of magnitudes more massive than the former, the above relation gives  $\< t_{ra} \> \sim N_0 m_{\rm max} m_{\rm min}^{4/5}$.
If both $m_{\rm min}$ and $m_{\rm max}$ are independent of $N_0$ then the relaxation time scales linearly in
with $N_0$ as in the standard case.

If, on the other hand,  one supposes that, because of increasing resolution,  $m_{\rm min} \sim  1/N_0$ (proportionately smaller
haloes appear in the resolved field as $N_0$ increases) then  $\< t_{ra} \> \sim  N_0^{1/5}$, which is roughly the scaling 
claimed by Diemand et al. (2004) on the basis of direct calculations of the relaxation time in the context of cosmological simulations.
Their results also show that the N-scaling is closer to the canonical linear relation as one moves towards the centre of the main halo. 
This is also to be expected; since, near the centre, stripping causes the fraction of mass in subhaloes to decrease, while the relaxation
in the smooth (parent halo) component becomes more efficient; and this scales roughly linearly with particle number.

Finally, note that although we have not followed the detailed merger history,
the expectation derived from the work presented here is that relaxation is more  enhanced  
in cluster haloes --- since these
form relatively recently, from particles that spend most of their existence inside poorly 
resolved structures. In contrast, a galaxy sized halo acquires most of its mass (i.e., particles) at larger redshift. 
 It follows that, given the same resolution at $z=0$,
cluster halo particles would have, on average, been more affected by discreteness noise. 
This is in line with the conclusion reached by Diemand et al. (2004).

%%%%%%%
\section*{ACKNOWLEDGMENTS}
I am indebted to
Milo\v s Milosavljevi\'c\\
for many valuable  discussions and suggestions, as well as 
critical comments on an earlier version of this paper 
that led to significant improvement.

\appendix
\section{Time dependence of $\rho_c$ and $z$ in flat universes with
cosmological constant}

According to Kolb \& Turner (P. 55), for a flat 
universe with cosmological constant one has
\begin{equation} t = \frac{2}{3} H^{-1} \Omega_V^{-1/2} \ln [
\frac{1+ \Omega_V^{1/2}}{(1- \Omega_V)^{1/2}}], 
\end{equation} 
which, using
$\Omega_V  = \frac{\Lambda c^2}{3 H^2}$ and $C_V = \sqrt{3 \Lambda} c$
gives
\begin{equation}
\Omega_V^{1/2} = \frac{ e^{C_V t} - 1}{ e^{C_V t} + 1}
\end{equation}
or equivalently
\begin{equation}
\frac{C_V}{3} \frac{ e^{C_V t} + 1}{ e^{C_V t} - 1} dt = - \frac{dz}{1+z},
\end{equation}
which by integration (with $A$ the associated constant) gives
\begin{equation}
1 + z = \frac{e^{\frac{C_V}{3}  t}} {A [e^{C_V t} -1]^{2/3}}.
\label{eq:zed}
\end{equation}
In a similar manner we find the critical density to vary as
\begin{equation}
\sqrt{\rho_C}= \frac{C_V}{2 \sqrt{6  \pi G}} \frac{e^{C_V t} + 1}{e^{C_V t} -1}.
\end{equation}
Using these formulas
the integral in Eq.~(\ref{eq:msdirect})
can be evaluated  with respect to  the variable $C_V t$. 
When the constant in Eq.~(\ref{eq:zed}) is evaluated by assuming $z=0$ at
$t=t_0$, the result does not depend on the value of $t_0$. This is analogous 
to the familiar case whereas two body relaxation in an isolated system 
does not depend on the time elapsed in physical units, but instead on the number
of dynamical times the system has been through.

\section{Velocity dispersions}

\begin{figure}
\psfig{file=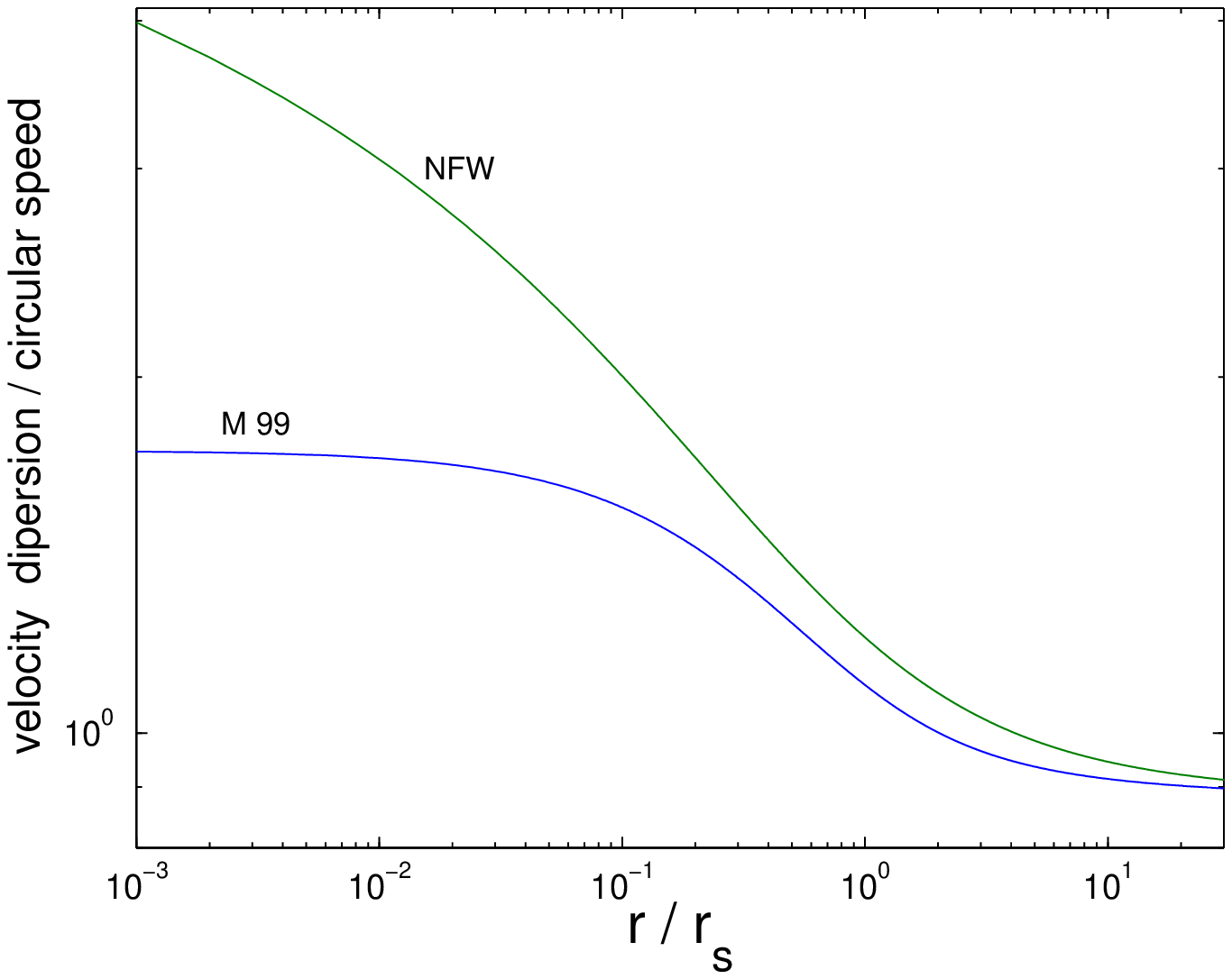,width=87mm}
%\onecolumn
\caption{Velocity dispersion in spherical cosmological haloes with 
isotropic velocities in terms of the local circular velocities as a 
function of radius scaled to the halos scale length $r_s$.}
\label{fig:velys}
\end{figure}

  The Jeans equation for spherical system with isotropic velocity dispersion can be written
as (e.g., Binney \& Tremaine 1987)
\begin{equation}
\frac{ d ( \rho \bar{ v_r^2 } ) } {dr} = - \rho \frac{d \Phi}{dr}, 
\end{equation}
where $\bar{v_r^2} = \frac{1}{3} \sigma^2$ is the radial velocity dispersion. 
The general solution is 
\begin{equation}
\rho  \bar{v_r^2} = - \int \rho \frac{d \Phi}{dr} dr + C.
\label{eq:velygen}
\end{equation}
For any $\rho \rightarrow 0$, as $r \rightarrow \infty$, one must have $C=0$, if
the velocity dispersion is to be bound at large radii.
The ratio of this unique4 physical solution  to the local circular velocity is
\begin{equation}
\frac{\bar{v_r^2}}{v_c^2} =  
\frac{\int_r^{\infty} \rho \frac{d \Phi}{dr} dr}{\rho \frac{d \Phi}{dr} r},
\end{equation}
the form reflecting the fact that, as opposed to circular motion,
 the velocity dispersion at any $r$ is caused by 
particles that venture in and out of that radius.

For power law density distributions $\rho \propto r^{-n}$
solutions of Eq.~(\ref{eq:velygen}), with $C=0$, imply that
\begin{equation}
\frac{\bar{v_r^2}}{v_c^2} = \frac{1}{2 n -2},
\end{equation}
for $n \ne 1$ and
\begin{equation}
\frac{\bar{v_r^2}}{v_c^2} = -  \ln r, 
\end{equation}
($r < 1$).
 Note that, in the central regions, $\bar{v_r^2} \propto - r \ln r$ for the 
NFW profile and $ \propto r^{1/2}$ for the M99 profile.
Spherical cosmological haloes thus, necessarily, have 
a `temperature inversion', at least as long  as the velocity dispersion can be considered isotropic. 

In Fig.~\ref{fig:velys} we reproduce the ratio of the three dimensional velocity dispersion 
to circular velocity for both profiles dealt with 
in this paper. It  determines  the value  of $gamma$ (corresponding to the average of this quantity 
within any given radius) that enter into the equations calculating the effects of relaxation.

\end{document}